\documentclass[aps,floats]{revtex4}
\usepackage{amsmath,amssymb}
\usepackage{graphicx,epsfig}

\begin{document}
\bibliographystyle {plain}

\def\oppropto{\mathop{\propto}} 
\def\opsimeq{\mathop{\simeq}}
\def\opoverderline{\mathop{\overline}}
\def\operarrow{\mathop{\longrightarrow}}
\def\opsim{\mathop{\sim}}

\def\fig#1#2{\includegraphics[height=#1]{#2}}
\def\figx#1#2{\includegraphics[width=#1]{#2}}


\title{ Dynamics of Ising models near zero temperature : \\
Real Space Renormalization Approach } 


 \author{ C\'ecile Monthus and Thomas Garel }
  \affiliation{ Institut de Physique Th\'{e}orique, CNRS and CEA Saclay,
 91191 Gif-sur-Yvette, France}

\begin{abstract}

We consider the stochastic dynamics of Ising ferromagnets (either pure or random) near zero temperature. The master equation satisfying detailed balance can be mapped onto a quantum Hamiltonian which has an exact zero-energy ground state representing the thermal equilibrium. The largest relaxation time $t_{eq}$ governing the convergence towards this Boltzmann equilibrium in finite-size systems is determined by the lowest non-vanishing eigenvalue $E_1=1/t_{eq}$ of the quantum Hamiltonian $H$. We introduce and study a real-space renormalization procedure for the quantum Hamiltonian associated to the single-spin-flip dynamics of Ising ferromagnets near zero temperature. We solve explicitly the renormalization flow for two cases. (i) For the one-dimensional random ferromagnetic chain with free boundary conditions, the largest relaxation time $t_{eq}$ can be expressed in terms of the set of random couplings for various choices of the dynamical transition rates. The validity of these RG results in $d=1$ is checked by comparison with another approach. (ii) For the pure Ising model on a Cayley tree of branching ratio $K$, we compute the exponential growth of $t_{eq}(N)$ with the number $N$ of generations.

\end{abstract}

\maketitle

\section{ Introduction }

The stochastic dynamics of classical Ising ferromagnets has been much
studied for fifty years \cite{lifshitz,glauber}.
In particular, many works have been devoted to the domain growth dynamics at low temperature $T<T_c$
(or at zero temperature $T=0$ in $d=1$ where the critical temperature vanishes $T_c=0$)
 when the initial condition is random (see the review on phase ordering dynamics \cite{bray}).

In the present paper, we do not consider the dynamics starting from a random initial condition,
but we focus instead on the largest relaxation time $t_{eq}$ needed for a finite
systems 
to converge towards thermal equilibrium. This largest relaxation time $t_{eq}$
 is defined as the inverse of the smallest non-vanishing
eigenvalue $E_1$ of the time-evolution operator.
Near zero-temperature, more precisely when the temperature is much smaller than any
ferromagnetic coupling $J_{ij}$
\begin{eqnarray}
0< T \ll J_{ij}
\label{lowT}
\end{eqnarray}
 the thermal equilibrium is dominated by the two ferromagnetic ground states where all spins
take the same value, and the largest relaxation time $t_{eq}$ corresponds to the time needed to go
from one ground state (where all spins take the value $+1$)
to the opposite ground state (where all spins take the value $-1$).
We should stress that we consider that the temperature is arbitrarily small, but does not
vanish, so that the transition between the two ground states is possible and the final state
of the dynamics is unique
(For studies on the zero-temperature dynamics, where the spin-flips corresponding to an energy-increase
become impossible, we refer to the recent works \cite{zeroT} and
to references therein).

Of course near zero temperature, the equilibration time $t_{eq}$ becomes extremely large,
and numerical simulations of
the microscopic dynamics become inefficient.
Here we introduce and study a real-space renormalization procedure
valid near zero temperature for the dynamics of pure or random ferromagnets.
The paper is organized as follows.
In section \ref{sec_hquantum}, we recall the standard mapping between the master equation
describing the stochastic dynamics
of classical  systems at temperature $T$ and a special type of quantum Hamiltonians that have an exact zero-energy eigenstate.
In section \ref{sec_singleflip}, we describe the various choices of dynamical transition
rates for single-spin-flip dynamics
of the classical Ising model, and the corresponding quantum Hamiltonians.
In section \ref{sec_rgrules}, we introduce the real-space renormalization procedure for
these general quantum Hamiltonians.
In section \ref{sec_brg1d}, we show how a closed RG procedure
 can be defined and exactly
solved for the random ferromagnetic chain.
In section \ref{sec_tree}, we solve the RG flow for the pure Ising model on a Cayley tree.
Section \ref{sec_conclusion} summarizes our conclusions.
In Appendix \ref{sec_app}, we describe another approach that allows to check the
validity of the RG results of section \ref{sec_brg1d}.
Finally in Appendix \ref{sec_diff}, we discuss the contributions
 that can depend on the choice of transition rates satisfying detailed balance.

\section{ Relaxation of classical models towards thermal equilibrium }

\label{sec_hquantum}

\subsection{ Master Equation satisfying detailed balance} 

To define the stochastic dynamics of a classical system,
it is convenient to consider the master equation 
\begin{eqnarray}
\frac{ dP_t \left({\cal C} \right) }{dt}
= \sum_{\cal C '} P_t \left({\cal C}' \right) 
W \left({\cal C}' \to  {\cal C}  \right) 
 -  P_t \left({\cal C} \right) W_{out} \left( {\cal C} \right)
\label{master}
\end{eqnarray}
that describes the time evolution of the
probability $P_t ({\cal C} ) $ to be in  configuration ${\cal C}$
 at time t.
The notation $ W \left({\cal C}' \to  {\cal C}  \right) $ 
represents the transition rate per unit time from configuration 
${\cal C}'$ to ${\cal C}$, and 
\begin{eqnarray}
W_{out} \left( {\cal C} \right)  \equiv
 \sum_{ {\cal C} '} W \left({\cal C} \to  {\cal C}' \right) 
\label{wcout}
\end{eqnarray}
represents the total exit rate out of configuration ${\cal C}$.

For a classical system where each configuration ${\cal C}$ has some energy
$U(C)$, the convergence towards Boltzmann equilibrium at temperature $T=\frac{1}{\beta}$
in any finite system
\begin{eqnarray}
P_{eq}({\cal C}) = \frac{ e^{- \beta U({\cal C})} }{Z}
\end{eqnarray}
where $Z$ is the partition function
\begin{eqnarray}
Z = \sum_{\cal C}  e^{- \beta U({\cal C})}
\label{partition}
\end{eqnarray}
can be ensured by imposing the detailed balance property
\begin{eqnarray}
e^{- \beta U({\cal C})}   W \left( \cal C \to \cal C '  \right)
= e^{- \beta U({\cal C '})}   W \left( \cal C' \to \cal C   \right)
\label{detailed}
\end{eqnarray}

\subsection{ Mapping onto a Schr\"odinger equation in configuration space}

As is well known (see for instance the textbooks \cite{gardiner,vankampen,risken}),
 the non-symmetric operator describing the stochastic  
dynamics of a classical model at temperature $T$
can be transformed into a symmetric quantum Hamiltonian problem.
In the field of disordered systems, this mapping has been much used for
one-dimensional continuous models  \cite{jpbreview,laloux,golosovloc,texier},
and more recently for many-body spin systems like the Sherrington-Kirkpatrik model
(\cite{us_conjugate} and Appendix B of \cite{castelnovo}).
In the field of pure spin models, 
this mapping has been used for more than fifty years
\cite{glauber,felderhof,siggia,kimball,peschel}.

In the present context, this standard mapping consists in 
the change of variable
\begin{eqnarray}
P_t ( {\cal C} ) \equiv e^{-  \frac{\beta}{2} U(\cal C ) } \psi_t ({\cal C} )
=   e^{-  \frac{\beta}{2} U(\cal C ) } <{\cal C} \vert  \psi_t  >
\label{relationPpsi}
\end{eqnarray}
Then the master equation of Eq. \ref{master}
becomes the imaginary-time  Schr\"odinger equation
for the ket  $\vert  \psi_t  >$ 
\begin{eqnarray}
\frac{ d }{dt} \vert  \psi_t  > = -H \vert  \psi_t  > 
\label{Hquantum}
\end{eqnarray}
where the quantum Hamiltonian 
\begin{eqnarray}
{\cal H} = \sum_{\cal C } \epsilon \left( {\cal C} \right) \vert {\cal C} > < {\cal C } \vert
+ \sum_{{\cal C},{\cal C '}}  V({\cal C} , {\cal C '})
 \vert {\cal C '} > < {\cal C } \vert
\label{tight}
\end{eqnarray}
contains the on-site energies 
\begin{eqnarray}
 \epsilon \left( {\cal C} \right) = W_{out} \left( {\cal C} \right)
\equiv
 \sum_{ {\cal C} '} W \left({\cal C} \to  {\cal C}' \right)
\label{eps}
\end{eqnarray}
and the hoppings (using Eq \ref{detailed})
\begin{eqnarray}
 V({\cal C } , {\cal C '}) && =- e^{- \frac{\beta}{2} \left[ U({\cal C '} )-U({\cal C }) \right] } W \left( \cal C' \to \cal C   \right)
= - e^{- \frac{\beta}{2} \left[ U({\cal C} )-U({\cal C '}) \right] } W \left( \cal C \to \cal C'   \right)
\nonumber \\
&& = - \sqrt {W \left( {\cal C} \to {\cal C '} \right) W \left( {\cal C '} \to {\cal C} \right) } 
\label{hopping}
\end{eqnarray}

\subsection{ Properties of the quantum Hamiltonian $\cal H$ }

Let us note $E_n$ the eigenvalues of the quantum Hamiltonian  ${\cal H}$ and $\vert \psi_n>$ the associated normalized eigenvectors
\begin{eqnarray}
{\cal H}  \vert  \psi_n > && = E_n \vert \psi_n> \\
\sum_{\cal C} \vert \psi_n({\cal C}) \vert ^2 && =1
\label{spectreH}
\end{eqnarray}
The evolution operator $e^{-t H}$ can be expanded on the eigenstates
\begin{eqnarray}
  e^{-t \cal H}  = 
 \sum_n e^{- E_n t} \vert \psi_n> < \psi_n\vert
\label{spectreHexp}
\end{eqnarray}

The conditional probability 
$P_t \left( {\cal C} \vert {\cal C}_0\right)$ to be in configuration ${\cal C}$ at $t$
if one starts from the configuration ${\cal C}_0$ at time $t=0$
can be written as
\begin{eqnarray}
P_t \left( {\cal C} \vert {\cal C}_0\right) =
   e^{-  \frac{\beta}{2} \left[U({\cal C} )-U({\cal C}_0 ) \right] } <{\cal C} \vert e^{-t\cal  H}  \vert {\cal C}_0>
= e^{-  \frac{\beta}{2} \left[U({\cal C} )-U({\cal C}_0 ) \right] }
\sum_n e^{- E_n t} \psi_n({\cal C})\psi_n^*({\cal C}_0)
\label{expansionP}
\end{eqnarray}

The quantum Hamiltonian $\cal H$ has special properties
that come from its relation to the dynamical master equation :

(i) the ground state energy is $E_0=0$, and the corresponding
eigenvector is given by

\begin{eqnarray}
\vert  \psi_0 > = \sum_{\cal C}  \frac{ e^{- \frac{\beta}{2} U({\cal C}) }}{\sqrt Z}
\vert { \cal C}  >
\label{psi0}
\end{eqnarray}
The normalization $1/\sqrt Z$ comes from the quantum normalization of Eq. \ref{spectreH}.

This property ensures the convergence towards the Boltzmann equilibrium
 in Eq. \ref{relationPpsi} for any initial condition ${\cal C}_0$
\begin{eqnarray}
P_t \left( {\cal C} \vert {\cal C}_0\right)
\opsimeq_{t \to + \infty}  e^{-  \frac{\beta}{2} \left[U({\cal C} )-U({\cal C}_0 ) \right]}
\psi_0({\cal C})\psi_0^*({\cal C}_0) = \frac{e^{- \beta U({\cal C})}}{Z} = P_{eq}({\cal C})
\label{CVeqP}
\end{eqnarray}

(ii) the other energies $E_n>0$ determine the relaxation towards equilibrium.
In particular, the lowest non-vanishing energy $E_1$
determines the largest relaxation time $(1/E_1)$ of the system 
\begin{eqnarray}
P_t \left( {\cal C} \vert {\cal C}_0\right) - P_{eq}({\cal C})
\opsimeq_{t \to + \infty} e^{- E_1 t}  e^{-  \frac{\beta}{2} \left[U({\cal C} )-U({\cal C}_0) \right] }
\psi_1({\cal C})\psi_1^*({\cal C}_0) 
\label{CVeqP1}
\end{eqnarray}
Since this largest relaxation time represents the 'equilibrium time', 
i.e. the characteristic time needed to converge towards equilibrium, 
we will use the following notation 
\begin{eqnarray}
t_{eq} \equiv \frac{1}{E_1}
\label{deftaueq}
\end{eqnarray}

In summary, the relaxation time $t_{eq}$ can be computed without simulating the dynamics
by any  method able
to compute the first excited energy $E_1$ of the quantum Hamiltonian $\cal H$
(where the ground state is given by Eq. \ref{psi0} and has for eigenvalue $E_0=0$). 
For instance in \cite{us_conjugate}, the 'conjugate gradient' method has been used
to study numerically the statistics of the largest relaxation time in various disordered models. Let us now describe more precisely how this general framework applies to single-spin-flip dynamics of Ising models.

\section{ Quantum Hamiltonian associated to single-spin-flip dynamics  } 

\label{sec_singleflip}

\subsection{ Single-spin-flip dynamics  } 

We consider a system of classical spins $S_i =\pm 1$ where each configuration ${\cal C}={S_1,S_2,..}$ has for energy
\begin{eqnarray}
U({\cal C}) = -\sum_{i<j} J_{ij} S_i S_j
\label{Uspin}
\end{eqnarray}
The couplings $J_{ij}$ may be random.

Within a single-spin flip dynamics, the configuration $\vert {\cal C} >= \vert S_1> \vert S_2 >..\vert S_N >$
containing $N$ spins
is connected via the transition rates $W \left( \cal C \to \cal C '  \right) $
to the $N$ configurations obtained by the flip of a single spin $S_k \to -S_k$
denoted by
\begin{eqnarray}
\vert {\cal C}_k > = \sigma^x_k \vert {\cal C} >
\label{ck}
\end{eqnarray}
in terms of the Pauli matrix $\sigma_x$.
The energy difference between the two configurations reads
\begin{eqnarray}
U({\cal C}_k)-U({\cal C}) =  2 S_k \sum_{i} J_{ki} S_i 
\label{Uspindiff}
\end{eqnarray}

The quantum Hamiltonian of Eqs \ref{tight} \ref{eps} \ref{hopping}
can be thus rewritten as
\begin{eqnarray}
{\cal H } && = \sum_{\cal C } \epsilon \left( {\cal C} \right) \vert {\cal C} > < {\cal C } \vert + \sum_{{\cal C}} \sum_{k=1}^N  V({\cal C} , \sigma^x_k {\cal C })
\sigma^x_k \vert {\cal C } > < {\cal C } \vert
\nonumber \\
 && =\sum_{\cal C } \sum_{k=1}^N  W \left({\cal C} \to  \sigma^x_k{\cal C} \right) 
e^{ \frac{\beta}{2} \left[ U({\cal C}_k )-U({\cal C }) \right]} 
\left[  e^{ - \frac{\beta}{2} \left[ U({\cal C}_k )-U({\cal C }) \right]} -  \sigma^x_k \right] \vert {\cal C } > < {\cal C } \vert
\label{Hreal}
\end{eqnarray}

\subsection{ Simplest choice  of the transition rates}

It is clear from Eq. \ref{Hreal} that the simplest quantum Hamiltonian
corresponds to the following choice of the dynamical transition rate 
\begin{eqnarray}
W \left( {\cal C}= \{ S_i \} \to {\cal C}_k = \sigma^x_k {\cal C}\right)
=    e^{-  \frac{\beta}{2} \left[ U({\cal C }_k)-U({\cal C }) \right] } 
= e^{ - \beta S_k \left( \sum_{i \ne k} J_{ik} S_i \right) }
\label{Wsimple}
\end{eqnarray}
The quantum Hamiltonian of Eq. \ref{Hreal} then reads in terms of Pauli matrices
\begin{eqnarray}
 {\cal H}^{simple}  &&
= \sum_{\cal C } \sum_{k=1}^N  
\left[  e^{ - \frac{\beta}{2} \left[ U({\cal C}_k )-U({\cal C }) \right]} -  \sigma^x_k \right] \vert {\cal C } > < {\cal C } \vert
\nonumber \\
&& 
= \sum_{\cal C } \sum_{k=1}^N  
\left[  e^{ - \beta  S_k \left[ \sum_{i} J_{ki} S_i  \right]} -  \sigma^x_k \right] 
\vert {\cal C } > < {\cal C } \vert
\nonumber \\
&& =  \sum_{ k=1 }^N \left[  e^{ - \beta \sigma^z_k \left( \sum_{i \ne k} J_{ik} \sigma^z_i \right) }
-   \sigma^x_k \right]
\label{Hsimple}
\end{eqnarray}
 where we have used the identity $1=\sum_{\cal C } \vert {\cal C } > < {\cal C } \vert$.
The quantum Hamiltonian of Eq. \ref{Hsimple}
has been mentioned as the simplest for the one-dimensional ferromagnetic chain in Eq (4) of 
Ref \cite{siggia}.

\subsection{ Glauber  choice  }
 
The Glauber choice for the transition rates \cite{glauber}
\begin{eqnarray}
W \left( {\cal C}= \{ S_i \} \to {\cal C}_k  = \sigma^x_k {\cal C}\right)
= \frac{   e^{-  \frac{\beta}{2} \left[ U({\cal C }_k)-U({\cal C }) \right] } }
{ 2 \cosh \left(\frac{\beta}{2} \left[ U({\cal C }_k)-U({\cal C }) \right] \right) }
= \frac{ e^{ - \beta S_k \left[ \sum_{i \ne k} J_{ik} S_i \right] }}
{ 2 \cosh \left( \beta  \left[ \sum_{i \ne k} J_{ik} S_i \right] \right) }
\label{glauber}
\end{eqnarray}
corresponds to the more complicated quantum Hamiltonian 
\begin{eqnarray}
 H^{Glauber} 
&& =  \sum_{ k=1 }^N  \frac{1}{2 \cosh \left[ \beta \left( \sum_{i \ne k} J_{ik} \sigma^z_i \right) \right]}  \left( e^{ - \beta \sigma^z_k \left( \sum_{i \ne k} J_{ik} \sigma^z_i \right) }
-   \sigma^x_k \right)
\label{Hglauber}
\end{eqnarray}
where we have used the fact that $ \sigma^z_k$ has for eigenvalues $(\pm1)$ and that $\cosh$
is an even function.

This quantum Hamiltonian of Eq. \ref{Hglauber}
has been used already used for the Sherrington-Kirkpatrick spin-glass model
and for the finite dimensional ferromagnetic Ising model
in Ref. \cite{castelnovo} (see Appendix B and Appendix C respectively).
For the one-dimensional pure ferromagnetic chain, where
each spin $S_k$ has only two neighbors, the local field $B_k \equiv \sum_{i } J_{ik} S_i
= J (S_{k-1}+S_{k+1})$ can take only the three values $h_k=-2 J, 0, 2J$, so that
 one may replace
the exponential factors using projection operators to recover the forms given in Refs \cite{siggia,kimball,peschel}.
This type of 'first quantized' quantum spin Hamiltonian can be transformed further into 'second quantized' Hamiltonian
involving annihilation/creation operators or Fermi operators using Jordan-Wigner transformation (see the review \cite{schutz}) :
this method has been followed in particular by Ref. \cite{felderhof} for the Glauber dynamics of the pure Ising chain.
However in the present paper, we will work directly on the 'first quantized' form of the quantum spin Hamiltonian
of Eq \ref{Hsimple} or Eq \ref{Hglauber}, with the aim to define a real-space renormalization approach,
in analogy with the Strong Disorder RG rules introduced for the random transverse field Ising model 
on its first-quantized form (see the discussion in section \ref{sec_strong} below).

\subsection{ Most general choice  }

To better understand the structure of the renormalized Hamiltonian
that will be generated by the real space RG procedure
introduced in the following section,
it is useful to consider
the most general choice satisfying the detailed balance equation of Eq. \ref{detailed}
\begin{eqnarray}
W \left( {\cal C}= \{ S_i \} \to {\cal C}_k  = \sigma^x_k {\cal C}\right)
= 
G_k(S_1,..,S_{k-1},S_{k+1},..,S_N)  e^{ - \beta S_k \left[ \sum_{i \ne k} J_{ik} S_i \right] }
\label{gene}
\end{eqnarray}
where $G_k(S_1,..,S_{k-1},S_{k+1},..,S_N) $ is an arbitrary positive function 
of the $(N-1)$ spins $i \ne k$ that may depend on the index $k$.
The positivity requirement 
\begin{eqnarray}
G_k(S_1,..,S_{k-1},S_{k+1},..,S_N)  >0
\label{Gkpositivity}
\end{eqnarray}
ensures that all elementary single-flip
are possible with a non-vanishing rate, so that the ground state of Eq. \ref{psi0}
is unique, and the dynamics converges towards thermal equilibrium.
Since the reversed transition of Eq. \ref{gene} has the following rate
\begin{eqnarray}
W \left( {\cal C}_k  = \sigma^x_k {\cal C} \to {\cal C}= \{ S_i \} \right)
= 
G_k(S_1,..,S_{k-1},S_{k+1},..,S_N)  e^{ + \beta S_k \left[ \sum_{i \ne k} J_{ik} S_i \right] }
\label{generev}
\end{eqnarray}
the amplitude $G_k(S_1,..,S_{k-1},S_{k+1},..,S_N)$ represents the 'symmetric part' of the two
opposite transitions involving the flip of the spin $k$ (when all other spins remain fixed)
\begin{eqnarray}
G_k(S_1,..,S_{k-1},S_{k+1},..,S_N)  = \sqrt {W \left( {\cal C} \to  \sigma^x_k {\cal C}\right) W \left( \sigma^x_k {\cal C} \to {\cal C} \right) } 
\label{Gsym}
\end{eqnarray}

The corresponding quantum Hamiltonian of Eq \ref{Hreal} reads
\begin{eqnarray}
{\cal H }^{general} && 
=  \sum_{ k=1 }^N  G_k(\sigma^z_1,..,\sigma^z_{k-1},\sigma^z_{k+1},..,\sigma^z_N)
  \left( e^{ - \beta \sigma^z_k \left( \sum_{i \ne k} J_{ik} \sigma^z_i \right) }
-   \sigma^x_k \right)
\label{Hgene}
\end{eqnarray}

In finite dimensions with only nearest-neighbors interactions,
it seems natural to consider local rates where the amplitude 
only involves the spins $i$ that are neighbors of $k$ (i.e. such that $J_{ki} \ne 0$)
\begin{eqnarray}
  G_k(\sigma^z_1,..,\sigma^z_{k-1},\sigma^z_{k+1},..,\sigma^z_N) 
\to  G_k^{local}( \{ \sigma^z_i \}_{J_{ik} \ne 0} \} )
\label{Glocalgene}
\end{eqnarray}
To respect the symmetries of the classical energy of Eq. \ref{Uspin},
a further requirement is usually that the amplitude should only 
be an even function $G(x)=G(-x)$ of the local field
\begin{eqnarray}
  G_k^{local}( \{ \sigma^z_i \}_{J_{ik} \ne 0} \} ) \to G \left( \sum_{i} J_{ki} \sigma^z_i\right)
\label{Flocalfield}
\end{eqnarray}
In the next section, we introduce a renormalization procedure 
for the amplitude $G_k$ and we will see that even if one starts with 
a function of the local field (Eq. \ref{Flocalfield}),
this form is not stable via renormalization and will generally lead
to a local function of the neighboring spins (Eq. \ref{Glocalgene}).

\section{ Renormalization rules near zero temperature }

\label{sec_rgrules}

In this section, we derive the appropriate renormalization rules
for the quantum Hamiltonian introduced in the previous section
in relation with the single-spin flip dynamics of classical ferromagnets,

\subsection{ Differences with the disordered quantum Ising model  }

\label{sec_strong}

Let us first mention that in the high-temperature limit $\beta J_{i,j} = \frac{J_{i,j}}{T} \ll  1$,
the quantum Hamiltonian of Eq. \ref{Hsimple} or \ref{Hglauber}
reduces to the standard transverse-field Ising model 
\begin{eqnarray}
 H^{simple,Glauber} 
\opsimeq_{\beta  J_{i,j} \ll 1 } N + \sum_{ k=1 }^N   \left(  - \beta \sigma^z_k \left( \sum_{i \ne k} J_{ik} \sigma^z_i \right) 
-   \sigma^x_k \right)
\label{HhighT}
\end{eqnarray}
When the coupling $J_{ij}$ are random, the low-energy physics is then well described
by the Strong Disorder RG procedure valid both in one dimension \cite{danieltransverse}
and in higher dimensions $d>1$ \cite{fisherreview,motrunich,kovacsreview} (see \cite{StrongRGreview} for a review).
However here we are interested into the opposite limit of very low temperature 
where $\beta J_{i,j} = \frac{J_{i,j}}{T} \gg  1 $ (Eq \ref{lowT}), where one cannot linearize the exponentials in the quantum Hamiltonians of Eq. \ref{Hsimple} and \ref{Hglauber}.
In the following, we thus derive appropriate RG rules in this opposite regime.

\subsection{ Analysis of an elementary operator }

The general quantum Hamiltonian of Eq \ref{Hgene} can be considered as a sum 
\begin{eqnarray}
 {\cal H}^{general}  =  \sum_{ k=1 }^N h_k 
\label{Hsimplesumlocal}
\end{eqnarray}
of elementary operators
\begin{eqnarray}
h_k \equiv G_k(\sigma^z_1,..,\sigma^z_{k-1},\sigma^z_{k+1},..,\sigma^z_N)
\left( e^{ - \beta \sigma^z_k \left( \sum_{i} J_{ki} \sigma^z_i \right) }
-   \sigma^x_k \right) 
\label{Hklocal}
\end{eqnarray}
The corresponding matrix elements in the $\sigma^z$ basis
\begin{eqnarray}
< S_1',..,S_N' \vert h_k \vert S_1,..,S_N > 
 = \left( \prod_{j \ne k} \delta_{S_j',S_j} \right) G_k(S_1,..,S_{k-1},S_{k+1},..,S_N)
< S_k' \vert  e^{ - \beta \sigma^z_k \left( \sum_{i  k} J_{ik} S_i \right) }
-   \sigma^x_k  \vert S_k >
\label{Hklocalmatrix}
\end{eqnarray}
are diagonal for all spins $j \ne k$.

\subsubsection{ Effective problem for a single spin } 

\label{elementaryoperator}

For each fixed value of the local field $B_k \equiv  \sum_{i } J_{ik} S_i $,
we may diagonalize the effective problem for the single-spin $k$
\begin{eqnarray}
h_k^{eff}(B_k) \equiv  e^{ - \beta \sigma^z_k B_k }-   \sigma^x_k  
\label{HklocalB}
\end{eqnarray}
The two normalized eigenstates are respectively
\begin{eqnarray}
\vert v_k(B_k) > && \equiv \frac{1}{ \sqrt{ 2 \cosh (\beta B_k  )  } }
\left[  e^{ \frac{\beta}{2} B_k } \vert S_k=+1 >
+  e^{ -\frac{\beta}{2} B_k  } \vert S_k=-1 > \right]
\label{vknormadef}
\end{eqnarray}
and 
\begin{eqnarray}
\vert w_k(B_k)> && \equiv 
\frac{1}{ \sqrt{ 2 \cosh (\beta B_k  )  } }
\left[  e^{ - \frac{\beta}{2} B_k } \vert S_k=+1 >
-  e^{ \frac{\beta}{2} B_k  } \vert S_k=-1 > \right]
\label{wknormadef}
\end{eqnarray}
with eigenvalues
\begin{eqnarray}
h_k^{eff}(B_k) \vert v_k(B_k) > && =0
\nonumber \\
 h_k^{eff}(B_k) \vert w_k(B_k) > && = 2 \cosh (\beta B_k  ) \vert w_k(B_k) >
\label{HklocalBeigen}
\end{eqnarray}
so that the single-spin hamiltonian of Eq. \ref{HklocalB} can be rewritten as the projector
\begin{eqnarray}
h_k^{eff}(B_k) && =  2 \cosh (\beta B_k  ) \vert  w_k(B_k) > <w_k(B_k) \vert 
\nonumber \\
= && \left[  e^{ - \frac{\beta}{2} B_k } \vert S_k=+1 >
-  e^{ \frac{\beta}{2} B_k  } \vert S_k=-1 > \right]
\left[  e^{ - \frac{\beta}{2} B_k } < S_k=+1 \vert
-  e^{ \frac{\beta}{2} B_k  } < S_k=-1 \vert \right]
\nonumber \\
= && \sigma_k^z  e^{- \frac{\beta}{2}  \sigma_k^z B_k  } 
\left[  \vert S_k=+1 >+ \vert S_k=-1 > \right]
\left[  < S_k=+1 \vert
+ < S_k=-1 \vert \right] \sigma_k^z  e^{- \frac{\beta}{2}  \sigma_k^z B_k  }
\label{HklocalBproj}
\end{eqnarray}

In terms of operators, the elementary operator of Eq. \ref{Hklocal} may be thus rewritten as
\begin{eqnarray}
h_k && = G_k(\sigma^z_1,..,\sigma^z_{k-1},\sigma^z_{k+1},..,\sigma^z_N)
\nonumber \\
&&  \sigma_k^z  e^{- \frac{\beta}{2}  \sigma_k^z \sum_i J_{ki} \sigma_i^z  } 
\left[  \vert S_k=+1 >+ \vert S_k=-1 > \right]
\left[  < S_k=+1 \vert
+ < S_k=-1 \vert \right] \sigma_k^z  e^{- \frac{\beta}{2}  \sigma_k^z \sum_i J_{ki} \sigma_i^z  }
\label{Hsimplesumlocalproj}
\end{eqnarray}

\subsubsection{ Properties of elementary operators } 

Eq \ref{Hsimplesumlocalproj} is convenient to see explicitly
 the positivity property for any ket $\vert \psi >$
\begin{eqnarray}
< \psi \vert h_k \vert  \psi> && = \sum_{S_1,..,S_{k-1},S_{k+1},..,S_N} 
G_k (S_1,..,S_{k-1},S_{k+1},..,S_N ) 
\nonumber \\
&& \vert e^{- \frac{\beta}{2} \sum_i J_{ki} S_i }
\psi(S_1,..,S_{k-1},S_k=+1, S_{k+1},..,S_N )
- e^{ \frac{\beta}{2} \sum_i J_{ki} S_i } 
\psi(S_1,..,S_{k-1},S_k=-1, S_{k+1},..,S_N )   \vert^2
\nonumber \\
&& \geq 0
\label{positivityhk}
\end{eqnarray}
as a consequence of the positivity of the $G_k$ (Eq \ref{Gkpositivity}).
Moreover, it is clear
 that the exactly known ground state of zero energy of Eq. \ref{psi0}
\begin{eqnarray}
\vert \psi_0> &&  = \sum_{S_1,..,S_N} <S_1,..S_N \vert \psi_0> \vert S_1,..,S_N >
= \frac{1}{\sqrt{ Z_N(\beta) }} \sum_{S_1,..,S_N} e^{ \frac{\beta}{2}
 \sum_{1 \leq i<j \leq N} J_{ij} S_i S_j } \vert  S_1,..,S_N >
\nonumber \\
 Z_N(\beta) && = \sum_{S_1,..S_N}  e^{ \beta \sum_{1 \leq i<j \leq N} J_{ij} S_i S_j }
\label{gsopearor}
\end{eqnarray}
satisfies
\begin{eqnarray}
\frac{\psi_0(S_1,..,S_{k-1},S_k=+1, S_{k+1},..,S_N )}{\psi_0(S_1,..,S_{k-1},S_k=-1, S_{k+1},..,S_N )} = e^{ \beta \sum_i J_{ki} S_i } 
\label{ratiopsi0}
\end{eqnarray}
and is thus annihilated by all elementary operators $h_k$ for $k=1,..,N$
\begin{eqnarray}
h_k \vert  \psi_0> =0
\label{hkannihilation}
\end{eqnarray}
as it should.

\subsection{ Renormalization of the sum of two neighboring elementary operators }

\subsubsection{ Sum of two neighboring elementary operators} 

The sum of two neighboring local operators of Eq. \ref{Hklocal} of index $k$ and $l$ with $J_{kl} \ne 0$ reads
\begin{eqnarray}
h_k +h_l  && =  G_k(\sigma^z_1,..,\sigma^z_{k-1},\sigma^z_{k+1},..,\sigma^z_N)
\left( e^{ - \beta \sigma^z_k \left(  J_{kl} \sigma^z_l + \sum_{i \ne l} J_{ki} \sigma^z_i \right) }
-   \sigma^x_k \right) 
\nonumber \\
&& + G_l(\sigma^z_1,..,\sigma^z_{l-1},\sigma^z_{l+1},..,\sigma^z_N)
\left( e^{ - \beta \sigma^z_l \left( J_{kl} \sigma^z_k + \sum_{j \ne k} J_{lj} \sigma^z_j \right) }
-   \sigma^x_l \right)
\label{Hklocalsum}
\end{eqnarray}
The corresponding matrix elements in the $\sigma^z$ basis
\begin{eqnarray}
< S_1',..,S_N' \vert h_k +h_l  \vert S_1,..,S_N > 
&&  = \left( \prod_{j \ne (k,l)} \delta_{S_j',S_j} \right) 
\nonumber \\
&&
[ G_k(S_1,..,S_{k-1},S_{k+1},..,S_N)
< S_k' \vert  e^{ - \beta \sigma^z_k \left(  J_{kl} S_l + \sum_{i \ne l} J_{ki} S_i \right) }
-   \sigma^x_k   \vert S_k >
\nonumber \\
&& + G_l(S_1,..,S_{l-1},S_{l+1},..,S_N)
< S_l' \vert e^{ - \beta \sigma^z_l \left( J_{kl} S_k + \sum_{i \ne k} J_{li} S_i \right) }
-   \sigma^x_l  \vert S_l > ]
\label{Hklocalmatrix2}
\end{eqnarray}
are diagonal for all spins $j \ne (k,l)$.

\subsubsection{ Effective two-spin problem } 

For each fixed value of all the other external spins $S_{j \ne (k,l)}$, using the notations
\begin{eqnarray}
 B_k && \equiv  \sum_{i \ne l} J_{ki} S_i  \nonumber \\
 B_l && \equiv  \sum_{j \ne k} J_{lj} S_j  
\label{defBkBl}
\end{eqnarray}
and 
\begin{eqnarray}
 g_k^{S_l}  && \equiv   G_k(S_1,..,S_{k-1},S_{k+1},.,S_l.,S_N) 
\nonumber \\
 g_l^{S_k} && \equiv  G_l(S_1,., S_k,..,S_{l-1},S_{l+1},..,S_N)
\label{defgkgl}
\end{eqnarray}
we have to diagonalize the effective problem for the two spins $(k,l)$
\begin{eqnarray}
h_{k,l}^{eff} \equiv 
   g_k^{\sigma^z_l}
\left( e^{ - \beta \sigma^z_k \left( J_{kl} \sigma^z_l +  B_k \right) }
-   \sigma^x_k \right)
 +  g_l^{\sigma^z_k} \left(  e^{ - \beta \sigma^z_l \left( J_{kl} \sigma^z_k + B_l \right) }
-   \sigma^x_l  \right)
\label{Hklocal2g}
\end{eqnarray}
The four-dimensional vector
\begin{eqnarray}
\vert u_{\lambda} > = \sum_{S_k=\pm,S_l=\pm} c_{\lambda}^{S_k,S_l} \vert S_k,S_l >
\label{vlambda}
\end{eqnarray}
 is an eigenvector of the operator $h_{k,l}^{eff} $ of Eq. \ref{Hklocal2g}
with eigenvalue $\lambda$ if
\begin{eqnarray}
0 && = \left[  e^{ - \beta  J_{kl}  }
\left( g_k^+ e^{ - \beta  B_k} + g_l^+ e^{ - \beta  B_l} \right) -\lambda \right]
 c_{\lambda}^{++}
 -  g_l^+ c_{\lambda}^{+-}- g_k^+ c_{\lambda}^{-+}
\nonumber \\
0 && = \left[  e^{ - \beta  J_{kl}  }
\left( g_k^- e^{  \beta  B_k} + g_l^- e^{  \beta  B_l} \right) -\lambda \right] c_{\lambda}^{--}
 -  g_k^- c_{\lambda}^{+-}- g_l^- c_{\lambda}^{-+}
\nonumber \\
0 && = \left[  e^{  \beta  J_{kl}  }
\left( g_k^- e^{ - \beta  B_k} + g_l^+ e^{  \beta  B_l} \right) -\lambda \right] c_{\lambda}^{+-}
 -  g_l^+ c_{\lambda}^{++}- g_k^- c_{\lambda}^{--}
\nonumber \\
0 && = \left[  e^{  \beta  J_{kl}  }
\left( g_k^+ e^{  \beta  B_k} + g_l^- e^{ - \beta  B_l} \right) -\lambda \right] c_{\lambda}^{-+}
 -   g_k^+ c_{\lambda}^{++}- g_l^- c_{\lambda}^{--}
\label{eigenHklocalB2g}
\end{eqnarray}

\subsubsection{ Finding the lowest non-zero eigenvalue $\lambda_1$ } 

We already know the exact ground state $\vert u_{\lambda=0} >$ of eigenvalue $\lambda=0$,
with components (not normalized here)
\begin{eqnarray}
c_{\lambda=0}^{S_k,S_l} = e^{  \frac{\beta}{2}  J_{kl}S_k S_l +  \frac{\beta}{2}  B_k S_k + \frac{\beta}{2}  B_l S_l} 
\label{gsHklocalB2g}
\end{eqnarray}
To find the exact three other eigenvalues, one has to solve the remaining cubic equation.
Here, since we are interested in the other small eigenvalue 
$ \lambda_1 \ll e^{  \beta  J_{kl}  } $, we will neglect $\lambda_1$ in the two last equations of Eq. \ref{eigenHklocalB2g}
to obtain
\begin{eqnarray}
  c_{\lambda}^{+-}
&& =  e^{  - \beta  J_{kl}  } \frac{ g_l^+ c_{\lambda}^{++} + g_k^- c_{\lambda}^{--} }
{g_k^- e^{ - \beta  B_k} + g_l^+ e^{  \beta  B_l}} 
\nonumber \\
   c_{\lambda}^{-+}
&& = e^{  - \beta  J_{kl}  } \frac{ g_k^+ c_{\lambda}^{++} + g_l^- c_{\lambda}^{--} }
{g_k^+ e^{  \beta  B_k} + g_l^- e^{ - \beta  B_l}}
\label{eigenHklocalB2glast}
\end{eqnarray}
 that we may replace in the two first equations of Eq. \ref{eigenHklocalB2g}
\begin{eqnarray}
0 && = \left[ 
 g_k^+ e^{ - \beta  B_k} + g_l^+ e^{ - \beta  B_l}  -\lambda_1 
 e^{ \beta  J_{kl}  }\right]
 c_{\lambda}^{++}
 -  g_l^+ \frac{ g_l^+ c_{\lambda}^{++} + g_k^- c_{\lambda}^{--} }
{g_k^- e^{ - \beta  B_k} + g_l^+ e^{  \beta  B_l}}
- g_k^+  \frac{ g_k^+ c_{\lambda}^{++} + g_l^- c_{\lambda}^{--} }
{g_k^+ e^{  \beta  B_k} + g_l^- e^{ - \beta  B_l}}
\nonumber \\
0 && = \left[  
 g_k^- e^{  \beta  B_k} + g_l^- e^{  \beta  B_l}  -\lambda_1  e^{  \beta  J_{kl}  }\right] c_{\lambda}^{--}
 -  g_k^-  \frac{ g_l^+ c_{\lambda}^{++} + g_k^- c_{\lambda}^{--} }
{g_k^- e^{ - \beta  B_k} + g_l^+ e^{  \beta  B_l}}
- g_l^-   \frac{ g_k^+ c_{\lambda}^{++} + g_l^- c_{\lambda}^{--} }
{g_k^+ e^{  \beta  B_k} + g_l^- e^{ - \beta  B_l}}
\label{eigenHklocalB2gfirst}
\end{eqnarray}

For $\lambda_1=0$, we recover the exact solution of Eq. \ref{gsHklocalB2g}
as it should.
The other eigenvalue reads
 \begin{eqnarray}
  \lambda_1  =  e^{-  \beta  J_{kl} } 2 \cosh[   \beta  (B_k  +   B_l) ]
 \frac{ g_k^+g_{k}^- ( g_l^+ e^{ \beta B_k } + g_l^- e^{- \beta B_k } ) 
+ g_l^+g_{l}^- ( g_k^+ e^{ \beta B_l } + g_k^- e^{- \beta B_l } )  }
{ g_k^+g_{k}^- +g_l^+g_{l}^- + g_k^+ g_l^+ e^{ \beta (B_k+B_l)} +g_k^- g_l^- e^{ - \beta (B_k+B_l)}  }
\label{lambda1gene}
\end{eqnarray}
with the corresponding components of the eigenvector $\vert u_{\lambda_1} >$ (not normalized here)
\begin{eqnarray}
c_{\lambda_1}^{++} && = e^{  \frac{\beta}{2}  J_{kl} -  \frac{\beta}{2}  B_k  - \frac{\beta}{2}  B_l } 
\nonumber \\
c_{\lambda_1}^{--} && = - e^{  \frac{\beta}{2}  J_{kl} + \frac{\beta}{2}  B_k  + \frac{\beta}{2}  B_l } 
\label{veclambda1g}
\end{eqnarray}
The two other components $c_{\lambda_1}^{+-} $ and $c_{\lambda_1}^{-+}$ are given by Eq \ref{eigenHklocalB2glast}.

\subsubsection{ Projection onto the two ferromagnetic states  } 

The projection of the operator of Eq. \ref{Hklocal2g}
onto its two lowest states of eigenvalues $\lambda_0=0$ et $\lambda_1$ 
reads
\begin{eqnarray}
h_{k,l}^{eff} \simeq \frac{\lambda_1}{ < u_{\lambda_1}\vert u_{\lambda_1} >}
  \vert u_{\lambda_1} > < u_{\lambda_1}\vert 
\label{HklocalB2lowest}
\end{eqnarray}

At the level of approximation we are working,
we wish to keep only the two ferromagnetic states 
$++$ and $--$ (the two other states $+-$ and $-+$ have been taken into account
in Eq. \ref{eigenHklocalB2glast} to produce renormalized rates
between $++$ and $--$ in Eq \ref{eigenHklocalB2gfirst}).
So we keep only the two following leading components in the eigenvector 
\begin{eqnarray}
  \vert u_{\lambda_1} >  \simeq e^{  \frac{\beta}{2}  J_{kl}}
 \left(e^{  -  \frac{\beta}{2}  (B_k  +  B_l) } \vert ++> 
- e^{   \frac{\beta}{2}  (B_k  +   B_l ) }  \vert -->  \right)
\label{vlambdaferrog}
\end{eqnarray}
with the corresponding normalization
\begin{eqnarray}
< u_{\lambda_1}  \vert u_{\lambda_1} >  \simeq e^{  \beta  J_{kl}} 
2 \cosh[   \beta  (B_k  +   B_l) ]
\label{normavlambdaferrog}
\end{eqnarray}
 Eq \ref{HklocalB2lowest} becomes
\begin{eqnarray}
h_{k,l}^{eff} \simeq && \frac{ \lambda_1 }
{ 2 \cosh[   \beta  (B_k  +   B_l) ] }
 \nonumber \\ && 
 \left(e^{  -  \frac{\beta}{2}  (B_k +  B_l) } \vert ++> 
- e^{   \frac{\beta}{2} ( B_k  +   B_l) }  \vert -->  \right)
 \left(e^{  -  \frac{\beta}{2}  (B_k  +  B_l) } < ++ \vert 
- e^{   \frac{\beta}{2}  (B_k  +   B_l) }  < -- \vert \right)
\label{HklocalB2lowestferrog}
\end{eqnarray}

So the two spins $S_k$ and $S_l$ now form a single renormalized ferromagnetic cluster,
that we may represent by a single spin with the two states
 \begin{eqnarray}
\vert + >_R && = \vert ++ > 
\nonumber \\
\vert - >_R && = \vert -- >
\label{spincluster}
\end{eqnarray}
with the renormalized external field (see Eq \ref{defBkBl})
 \begin{eqnarray}
B_R=  B_k+B_l  && =  \sum_{i \ne (k,l)} (J_{ki}+J_{li}) S_i  
\label{Br}
\end{eqnarray}
that corresponds to the natural renormalization of ferromagnetic coupling
between the new cluster $(kl)$ and $i$
 \begin{eqnarray}
J^R_{(kl)i } = J_{ki}+J_{li} 
\label{Jr}
\end{eqnarray}
In terms of this renormalized cluster, Eq. \ref{HklocalB2lowestferrog} reads
\begin{eqnarray}
h_{k,l}^{eff} \simeq && G_R  \left(e^{  -  \frac{\beta}{2} B_R } \vert +>_R 
- e^{   \frac{\beta}{2} B_R }  \vert ->_R  \right)
 \left(e^{  -  \frac{\beta}{2} B_R } < + \vert_R 
- e^{   \frac{\beta}{2} B_R }  < - \vert_R \right)
\label{HklocalB2lowestferrorgg}
\end{eqnarray}
It has the same form as an elementary operator of Eq. \ref{Hsimplesumlocalproj}
with the renormalized amplitude (using Eq. \ref{lambda1gene})
 \begin{eqnarray}
G_R &&  =\frac{ \lambda_1 }
{ 2 \cosh[   \beta  (B_k  +   B_l) ]  }
\nonumber \\
&& =  e^{-  \beta  J_{kl} } 
 \frac{ g_k^+g_k^- ( g_l^+ e^{ \beta B_k } + g_l^- e^{- \beta B_k } ) 
+ g_l^+g_l^- ( g_k^+ e^{ \beta B_l } + g_k^- e^{- \beta B_l } )  }
{ g_k^+g_k^- +g_l^+g_l^- + g_k^+ g_l^+ e^{ \beta B_R} +g_k^- g_l^- e^{ - \beta B_R}  }
\label{ggene}
\end{eqnarray}
in terms of the variables defined in Eq. \ref{defgkgl}
for a given value of the external spins $S_{i \ne (k,l)}$.

\subsubsection{ Renormalization rules for operators } 

At the level of operators in the whole Hilbert space,
the renormalized Hamiltonian of Eq. \ref{HklocalB2lowestferrog} reads
(using the same notations as in Eq. \ref{Hklocal} )
\begin{eqnarray}
h_{(k,l)}^R \equiv G_{kl}^R(\sigma^z_1,..,\sigma^z_{k-1},\sigma^z_{k+1},.\sigma^z_{l-1},\sigma^z_{l+1}.,\sigma^z_N)
\left( e^{ - \beta \sigma^z_R B_R }
-   \sigma^x_R \right) 
\label{Hrgop}
\end{eqnarray}
where the amplitude of Eq. \ref{ggene} is the renormalized operator
 \begin{eqnarray}
G_{kl}^R (\sigma^z_1,..,\sigma^z_{k-1},\sigma^z_{k+1},.\sigma^z_{l-1},\sigma^z_{l+1}.,\sigma^z_N)
&& =  e^{-  \beta  J_{kl} } 
 \frac{ g_k^+g_k^- ( g_l^+ e^{ \beta B_k } + g_l^- e^{- \beta B_k } ) 
+ g_l^+g_l^- ( g_k^+ e^{ \beta B_l } + g_k^- e^{- \beta B_l } )  }
{ g_k^+g_k^- +g_l^+g_l^- + g_k^+ g_l^+
 e^{ \beta (B_k  +   B_l)} +g_k^- g_l^- e^{ - \beta (B_k  +   B_l)}  }
\label{ggeneop}
\end{eqnarray}
in terms of the operators
\begin{eqnarray}
 g_k^{\pm}  && \equiv   G_k(\sigma^z_1,..,\sigma^z_{k-1},\sigma^z_{k+1},.,S_l=\pm,.,\sigma^z_N) 
\nonumber \\
 g_l^{\pm} && \equiv  G_l(\sigma^z_1,., S_k=\pm,..,\sigma^z_{l-1},\sigma^z_{l+1},..,\sigma^z_N)
\nonumber \\
 B_k && \equiv  \sum_{i \ne l} J_{ki} \sigma^z_i  
\nonumber \\
 B_l && \equiv  \sum_{i \ne k} J_{li} \sigma^z_i  
\label{defgkglop}
\end{eqnarray}

\subsubsection{ Final State of the RG procedure } 

For a system of $N$ spins, we will obtain after $(N-1)$ RG steps a single renormalized spin $S_R=\pm 1$
representing the two ground states of the whole sample, where all spins take the same value $S_R$.
Since there is no renormalized external field left $B_R=0$, the final effective Hamiltonian 
for the single spin $S_R$ simply reads
\begin{eqnarray}
{\cal H}_N^{final} =  G^{final}(N) \left(1- \sigma_R^x \right)
\label{Gunity}
\end{eqnarray}
where $G^{final} (N)$ is a numerical amplitude.
The quantum 
ground state of zero energy is the symmetric combination of the two classical 
ferromagnetic ground states as it should
\begin{eqnarray}
\vert \psi_0^{(final)} > = \frac{\vert S_R=+ >+\vert S_R=- >}{\sqrt 2}
\label{psizerofinal}
\end{eqnarray}
whereas the excited quantum eigenstate is the antisymmetric combination of the two classical 
ferromagnetic ground states
\begin{eqnarray}
\vert \psi_1^{(final)} > = \frac{\vert S_R=+ >-\vert S_R=- >}{\sqrt 2}
\label{psi1final}
\end{eqnarray}
with the eigenvalue
\begin{eqnarray}
E_1^{final}(N) =  2 G^{final}(N) 
\label{E1finalG}
\end{eqnarray}
The conclusion is that the equilibrium time of Eq. \ref{deftaueq} can be thus obtained from the final renormalization amplitude $G_{N}^{final} $ as 
\begin{eqnarray}
t_{eq}(N) \equiv \frac{1}{E_1^{final}(N)} = \frac{1}{ 2 G^{final}(N)}
\label{taueqGfinal}
\end{eqnarray}
In summary, we have thus defined a renormalized procedure for the amplitudes $G$
that allows to obtain in the end the equilibration time $t_{eq}(N)$.
Now to better understand the meaning of the RG rules of Eq. \ref{ggeneop},
let us give some explicit examples.

\subsubsection{ Example : first renormalization step starting from the simple Hamiltonian of Eq. \ref{Hsimple} } 

The quantum Hamiltonian of Eq. \ref{Hsimple} corresponds to the initial
simple case where all amplitudes are unity
\begin{eqnarray}
 G_{k}^{ini}(\sigma^z_1,..,\sigma^z_{k-1},\sigma^z_{k+1},..,\sigma^z_N) =1
\label{Giniunity}
\end{eqnarray}
The first RG step where $k$ and $l$ are grouped into a single renormalized cluster
yields the following amplitude (Eq \ref{ggeneop} with $ g_k^{\pm}=1= g_l^{\pm} $ 
 \begin{eqnarray}
G_{kl}^R 
&& =  e^{-  \beta  J_{kl} } 
 \frac{  \cosh( \beta B_k ) +   \cosh( \beta B_l )  }
{ 1 +  \cosh( \beta (B_k+B_l) ) }
\nonumber \\
&& =  e^{-  \beta  J_{kl} } 
 \frac{  \cosh( \beta \sum_{i \ne l} J_{ki} \sigma^z_i  ) 
+   \cosh( \beta \sum_{i \ne k} J_{li} \sigma^z_i )  }
{ 1 +  \cosh( \beta \sum_{i \ne (k,l)} (J_{ki}+J_{li}) \sigma^z_i ) }
\label{ggeneopsimplefirststep}
\end{eqnarray}
So this renormalized amplitude does not remain a number as in the initial condition of
Eq \ref{Gunity}, but becomes an operator that involves the neighboring spins of $k$ and $l$.

\subsubsection{ Example : first renormalization step starting from the Glauber Hamiltonian of Eq. \ref{Hglauber} } 

The quantum Hamiltonian of Eq. \ref{Hglauber} corresponds to the initial
 case 
\begin{eqnarray}
 G_{k}^{ini}(\sigma^z_1,..,\sigma^z_{k-1},\sigma^z_{k+1},..,\sigma^z_N) =
 \frac{1}{2 \cosh (\beta \sum_{i } J_{ki} \sigma^z_i )}
\label{Gglauber}
\end{eqnarray}
so that the operators of Eq. \ref{defgkglop} reads
 \begin{eqnarray}
 g_{k}^{\pm} && = \frac{1}{2 \cosh \beta (B_k \pm J_{kl}  ) }
\nonumber \\
 g_{l}^{\pm} && = \frac{1}{2 \cosh \beta ( B_l \pm J_{kl}) }
\label{g12glauber}
\end{eqnarray}
Eq \ref{ggeneop} for the renormalized amplitude then becomes
 \begin{eqnarray}
G_{kl}^R 
&& =  \frac{1}{ 2 \cosh \beta (B_k+B_l) +2 e^{ 2 J_{kl}+B_k-B_l}}
+  \frac{1}{ 2 \cosh \beta (B_k+B_l) +2 e^{ 2 J_{kl}-B_k+B_l}}
\nonumber \\
&&=  \frac{1}{ 2 \cosh (\beta \sum_{i \ne (k,l)} (J_{ki}+J_{li}) \sigma^z_i )
 +2 e^{ 2 J_{kl}+\sum_{i \ne (k,l)} (J_{ki}-J_{li}) \sigma^z_i )}}
\nonumber \\
&& +  \frac{1}{ 2 \cosh (\beta \sum_{i \ne (k,l)} (J_{ki}+J_{li}) \sigma^z_i ) +2 e^{ 2 J_{kl}  - \sum_{i \ne (k,l)} (J_{ki}-J_{li}) \sigma^z_i )}}
\label{ggeneopglauberfirststep}
\end{eqnarray}
So this renormalized amplitude does not remain of the Glauber form of Eq. \ref{Gglauber}
  in terms of the renormalized local field $B_R = B_k+B_l$. Actually it does not even remain
a function of the single renormalized local field $B_R $.

\subsubsection{ Discussion  }

These two examples show that the renormalized amplitudes $G_R$ become generically a function of the $\sigma^z$ operators of the neighboring spins, ie of the form of Eq. \ref{Glocalgene}.
It is not clear to us at this stage how to determine the operator form that would
remain stable upon the general RG rule of Eq. \ref{ggeneop}.
In the following, we will thus concentrate on two geometries where we can obtain 
closed RG rules, namely the one-dimensional case and the Cayley tree.

\section{ Boundary Renormalization for the random ferromagnetic chain}

\label{sec_brg1d}

In this section, we consider the random ferromagnetic chain of $N$ spins
with the classical energy (Eq \ref{Uspin})
\begin{eqnarray}
U(S_1,...S_N) = -\sum_{i=1}^{N-1} J_{i,i+1} S_i S_{i+1}
\label{Uspin1d}
\end{eqnarray}
with free boundary conditions for the two boundary spins $S_1$ and $S_N$.
The couplings $J_{i,i+1} $ are positive random variables.

\subsection{ Closed RG for the simple Hamiltonian of Eq \ref{Hsimple} in one dimension   }

\subsubsection{ First step of the RG rules in the bulk   }

The first RG step where the two neighboring sites $(k,k+1)$ are grouped into a single
ferromagnetic cluster leads to the following renormalized amplitude (Eq \ref{ggeneopsimplefirststep}) for $1<k<N-1$
 \begin{eqnarray}
G_{k,k+1}^R 
&& =  e^{-  \beta  J_{k,k+1} } 
 \frac{  \cosh( \beta  J_{k-1,k}   ) 
+   \cosh( \beta  J_{k+1,k+2}  )  }
{ 1 +  \cosh \left( \beta \left[ J_{k-1,k} \sigma^z_{k-1} + J_{k+1,k+2} \sigma^z_{k+2} \right] \right) }
\label{ggeneopsimplefirststep1d}
\end{eqnarray}
Since there remains the renormalized local field operator $\left[ J_{k-1,k} \sigma^z_{k-1} + J_{k+1,k+2} \sigma^z_{k+2} \right]$
in the denominator, this renormalized amplitude is not just a number as the initial condition.
To avoid this difficulty,
we will now consider what happens near one boundary.

\subsubsection{ First step of the RG rules near the boundary   }

We now consider the case $k=1$, where we make a ferromagnetic cluster out of the
two sites $(1,2)$ near the boundary.
Since the spin zero does not exist, we have $J_{k-1,k}=J_{0,1}=0$
, so that Eq. \ref{ggeneopsimplefirststep1d} reduces to
a renormalized number (without any operator anymore)
 \begin{eqnarray}
G_{1,2}^R 
&& =  e^{-  \beta  J_{1,2} } 
 \frac{  1
+   \cosh( \beta  J_{2,3}  )  }
{ 1 +  \cosh ( \beta J_{2,3} ) } =  e^{-  \beta  J_{1,2} } 
\label{ggeneopsimplefirststep1db}
\end{eqnarray}
This shows that we may define a simple closed boundary RG procedure as follows.

\subsubsection{ Closed Boundary RG procedure   }

\label{closedsimple}

If we iterate the renormalization near the boundary,
the renormalized state after $(n-1)$ RG steps will be the following :
the spins $(1,2,..,n)$ have been grouped together into a single renormalized spin
with some associated renormalized amplitude $ G_{[1,n]}^R$.
The other spins $(n+1,...,N)$ of the chain are still in their initial form
with amplitude unity $G_k=1$ for $n+1 \leq k \leq N $.
Let us now perform the $n$ RG step where the spin $(n+1)$ is included in the boundary cluster. The RG rule of Eq. \ref{ggeneop} yields the recurrence
 \begin{eqnarray}
G_{[1,n+1]}^R
&& =  e^{-  \beta  J_{n,n+1} } G_{[1,n]}^R  \
\left[  \frac{  2 G_{[1,n]}^R +  2 \cosh (\beta J_{n+1,n+2})   }
{ (G_{[1,n]}^R)^2 +1 + G_{[1,n]}^R ( 2 \cosh (\beta J_{n+1,n+2}))    } \right]
\label{ggeneopclosedsimple}
\end{eqnarray}

Within the {\it bulk}, i.e. for $ J_{n+1,n+2} >0$, in order to 
 be consistent with our previous approximations in the low-temperature limit
(Eq \ref{lowT}), we may replace
 \begin{eqnarray}
 2 \cosh (\beta J_{n+1,n+2}) ) \opsimeq_{ \beta J_{n+1,n+2} \gg 1} e^{ \beta J_{n+1,n+2}}
\label{coshapprox}
\end{eqnarray}
Moreover, we expect that the renormalized amplitudes $G^R$ only decay upon renormalization, and are thus smaller than their initial unity value
(see already the first step of Eq. \ref{ggeneopsimplefirststep1db})
 \begin{eqnarray}
G_{[1,n]}^R \leq 1 
\label{grsmall}
\end{eqnarray}
So Eq. \ref{ggeneopclosedsimple} reduces to
 \begin{eqnarray}
G_{[1,n+1]}^R
&& \simeq  e^{-  \beta  J_{n,n+1} } G_{[1,n]}^R \ 
\left[ \frac{    e^{ \beta J_{n+1,n+2}}   }
{  1 + G_{[1,n]}^R  e^{ \beta J_{n+1,n+2}}  } \right]
\label{ggeneopclosedsimpleapp}
\end{eqnarray}
i.e. this recurrence becomes simpler in terms of inverse variables
 \begin{eqnarray}
\frac{e^{   \beta J_{n+1,n+2}}}{G_{[1,n+1]}^R}
&& \simeq   e^{ \beta  J_{n,n+1} + \beta J_{n+1,n+2}}
+  \frac{ e^{  \beta  J_{n,n+1} }   }{    G_{[1,n]}^R}  
\label{ginvsimple}
\end{eqnarray}

\subsubsection{ Final result for a chain of $N$ spins  }

For a finite chain of $N$ spins where $J_{N,N+1}=0$, the recurrence of Eq. \ref{ggeneopclosedsimple}
yields for the last step with $n+1=N$
 \begin{eqnarray}
G_{[1,N]}^{Rfinal}
&& =  e^{-  \beta  J_{N-1,N} } G_{[1,N-1]}^R  \
\left[  \frac{  2 G_{[1,N-1]}^R +  2   }
{ (G_{[1,N-1]}^R)^2 +1 + G_{[1,N-1]}^R  2     } \right]  
\nonumber \\
&&
= e^{-  \beta  J_{N-1,N} } G_{[1,N-1]}^R  \
\left[  \frac{  2   }
{ 1 + G_{[1,N-1]}^R       } \right]
\label{ggeneopclosedsimplefinal}
\end{eqnarray}
i.e. using inverse variables
 \begin{eqnarray}
\frac{1}{G_{[1,N]}^{Rfinal}}
= \frac{1}{2}
\left[   e^{  \beta  J_{N-1,N} }      + \frac{ e^{  \beta  J_{N-1,N} }}{ G_{[1,N-1]}^R} \right]
\label{simplefinalinverse}
\end{eqnarray}

We now now use iteratively the recurrence of Eq. \ref{ginvsimple} valid in
the bulk to obtain
 \begin{eqnarray}
\frac{1}{G_{[1,N]}^R} 
&&  =  \frac{1}{2}
\left[   e^{  \beta  J_{N-1,N} }
+   e^{ \beta  J_{N-2,N-1} + \beta J_{N-1,N}}
+  \frac{ e^{  \beta  J_{N-2,N-1} }   }{    G_{[1,N-2]}^R}  
 \right]
\nonumber \\
&&  =  \frac{1}{2}
\left[   e^{  \beta  J_{N-1,N} }
+   e^{ \beta  J_{N-2,N-1} + \beta J_{N-1,N}}
+  e^{ \beta  J_{N-3,N-2} + \beta J_{N-2,N-1}}
+  \frac{ e^{  \beta  J_{N-3,N-2} }   }{    G_{[1,N-3]}^R} 
 \right]
\nonumber \\
&& \simeq  \frac{1}{2} \sum_{k=1}^N e^{ \beta (J_{k-1,k}+J_{k,k+1}) }
\label{gfinalsimple}
\end{eqnarray}
(with the notations $J_{0,1}=0=J_{N,N+1}$)

The conclusion of this RG procedure is thus
that the equilibration time of the finite random chain of $N$ spins
reads for the 'simple' dynamics (Eq. \ref{taueqGfinal})
\begin{eqnarray}
t_{eq}^{simple}(N) = \frac{1}{E_1^{final}(N)} = \frac{1}{ 2G_{[1,N]}^R }
= \frac{1}{ 4 }  \sum_{k=1}^N e^{ \beta (J_{k-1,k}+J_{k,k+1}) }
\label{taueqsimple}
\end{eqnarray}

\subsection{ Closed RG for dynamics depending only on the local field in one dimension   }

We now consider the more general case where the amplitude $G_k$ are a single even function $G(x)=G(-x)$ of the local field (see Eq. \ref{Flocalfield}).
\begin{eqnarray}
 G_k =  G \left( J_{k-1,k} \sigma^z_{k-1} + J_{k,k+1} \sigma^z_{k+1} \right)
\label{Flocalfield1d}
\end{eqnarray}
The Glauber Hamiltonian of Eq \ref{Hglauber} corresponds to the special function
\begin{eqnarray}
   G (x) = \frac{1}{2 \cosh ( \beta x)}
\label{Gglaubercosh}
\end{eqnarray}

\subsubsection{ Closed Boundary RG procedure   }

If we iterate the renormalization near the boundary as in section \ref{closedsimple}, we obtain finally the following recurrence (Eq \ref{ggeneop}) :

 \begin{eqnarray}
G_{[1,n+1]}^R 
&& =  e^{-  \beta  J_{n,n+1} } G_{[1,n]}^R 
 \frac{ G_{[1,n]}^R  ( f_{n+1}^+  + f_{n+1}^-  ) 
+ f_{n+1}^+f_{n+1}^-   ( 2 \cosh (  \beta J_{n+1,n+2}))   }
{ ( G_{[1,n]}^R )^2+f_{n+1}^+f_{n+1}^- + G_{[1,n]}^R ( f_{n+1}^+
 e^{ \beta J_{n+1,n+2}} + f_{n+1}^- e^{ - \beta J_{n+1,n+2} } ) }
\label{ggeneoprecloc}
\end{eqnarray}
in terms of the numbers
\begin{eqnarray}
 f_{n+1}^{\pm} && \equiv   G \left(J_{n,n+1} \pm J_{n+1,n+2} \right)
\label{fpm}
\end{eqnarray}

Within the {\it bulk}, i.e. for $ J_{n+1,n+2} >0$, in order to 
 be consistent with our previous approximations in the low-temperature limit
(Eq \ref{lowT}), we may use Eqs \ref{coshapprox} and the fact that the renormalized amplitudes $G^R$ only decay upon renormalization to simplify Eq \ref{ggeneoprecloc}
into
 \begin{eqnarray}
G_{[1,n+1]}^R
&& \simeq  
  e^{-  \beta  J_{n,n+1} } G_{[1,n]}^R 
 \frac{ 
 f_{n+1}^+f_{n+1}^-   e^{  \beta J_{n+1,n+2} }  }
{ f_{n+1}^+f_{n+1}^- + G_{[1,n]}^R  f_{n+1}^+ e^{ \beta J_{n+1,n+2}}
 }
\label{ggeneopclosedgeneapp}
\end{eqnarray}
i.e. this recurrence becomes simpler in terms of inverse variables
 \begin{eqnarray}
\frac{e^{   \beta J_{n+1,n+2}}}{G_{[1,n+1]}^R}
&& \simeq  
 \frac{  e^{  \beta  J_{n,n+1} + \beta J_{n+1,n+2}} }
{f_{n+1}^-    }
+
 \frac{   e^{  \beta  J_{n,n+1} }  }
{ G_{[1,n]}^R     }
\label{ginvgene}
\end{eqnarray}

\subsubsection{ Final result for a chain of $N$ spins  }

For a finite chain of $N$ spins where $J_{N,N+1}=0$, 
 Eq \ref{fpm} reads
\begin{eqnarray}
 f_{N}^{\pm} && \equiv   G \left(J_{N-1,N} \pm 0 \right) =  f_{N}^{-}
\label{fpmend}
\end{eqnarray}
and the recurrence of Eq. \ref{ggeneoprecloc}
yields for the last step with $n+1=N$
 \begin{eqnarray}
G_{[1,N]}^{Rfinal}
&& =   e^{-  \beta  J_{N-1,N} } G_{[1,N-1]}^R 
 \frac{  2  f_{N}^- (G_{[1,N-1]}^R+f_{N}^- )  
   }
{ ( G_{[1,N-1]}^R +f_{N}^-)^2 }
\nonumber \\
&&
= e^{-  \beta  J_{N-1,N} } G_{[1,N-1]}^R  \
  \frac{  2 f_{N}^-  }
{  G_{[1,N-1]}^R  +f_{N}^-      } 
\label{ggeneopclosedgenefinal}
\end{eqnarray}
i.e. using inverse variables
 \begin{eqnarray}
\frac{1}{G_{[1,N]}^{Rfinal}}
= \frac{1}{2}
\left[  \frac{ e^{  \beta  J_{N-1,N} }}{f_{N}^-}       + \frac{ e^{  \beta  J_{N-1,N} }}{ G_{[1,N-1]}^R} \right]
\label{genefinalinverse}
\end{eqnarray}

We now now use iteratively the recurrence of Eq. \ref{ginvgene} valid in
the bulk to obtain
 \begin{eqnarray}
\frac{1}{G_{[1,N]}^R} 
&&  =  \frac{1}{2}
\left[   e^{  \beta  J_{N-1,N} }
+  \frac{ e^{ \beta  J_{N-2,N-1} + \beta J_{N-1,N}}}{f_{N-1}^-}
+  \frac{ e^{  \beta  J_{N-2,N-1} }   }{    G_{[1,N-2]}^R}  
 \right]
\nonumber \\
&&  =  \frac{1}{2}
\left[  \frac{ e^{  \beta  J_{N-1,N} } }{f_{N}^-}
+  \frac{ e^{ \beta  J_{N-2,N-1} + \beta J_{N-1,N}}}{f_{N-1}^-}
+ \frac{ e^{ \beta  J_{N-3,N-2} + \beta J_{N-2,N-1}}}{f_{N-2}^-}
+  \frac{ e^{  \beta  J_{N-3,N-2} }   }{    G_{[1,N-3]}^R} 
 \right]
\nonumber \\
&& \simeq  \frac{1}{2} \sum_{k=1}^N \frac{ e^{ \beta (J_{k-1,k}+J_{k,k+1}) } }{f_{k}^-}
\label{gfinalgene}
\end{eqnarray}
(with the notations $J_{0,1}=0=J_{N,N+1}$)

Using Eq. \ref{fpm}, the conclusion of this RG procedure is thus
that the equilibration time of the finite random chain of $N$ spins
reads for the dynamics defined by the amplitudes of Eq. \ref{ggeneoprecloc}
\begin{eqnarray}
t_{eq}(N) = \frac{1}{E_1^{final}(N)} = \frac{1}{ 2G_{[1,N]}^R }
= \frac{1}{ 4 }  \sum_{k=1}^N \frac{e^{ \beta (J_{k-1,k}+J_{k,k+1}) }}{G \left(J_{k-1,k} - J_{k,k+1} \right)}
\label{taueqgene}
\end{eqnarray}

In particular for the Glauber dynamics corresponding to Eq. \ref{Gglaubercosh},
the equilibration time reads
\begin{eqnarray}
t_{eq}^{glauber}(N)
&&  =  \frac{1}{ 4 }
  \sum_{k=1}^N e^{ \beta (J_{k-1,k}+J_{k,k+1}) }
2 \cosh \left( \beta J_{k-1,k} - \beta J_{k,k+1} \right)
 =  \frac{1}{ 4 }
  \sum_{k=1}^N 
\left[ e^{2 \beta J_{k-1,k}  } + e^{ 2 \beta  J_{k,k+1} }  \right]
\nonumber \\
&&  =  \frac{1}{ 2 }
\left[ 1 +  \sum_{k=1}^{N-1}  e^{2 \beta J_{k,k+1}  }  \right]
\label{taueqglauber}
\end{eqnarray}

\subsection{ Discussion  }

In this section, we have obtained a closed RG procedure in dimension $d=1$
that yields the explicit result of Eq. \ref{taueqgene} for the equilibration time
of a finite chain. The validity of this RG result is checked in Appendix \ref{sec_app}
using another method. Since we compare the explicit expression in terms of the set of all random couplings,
for various choices of the transition rates,
this agreement shows the exactness of the RG procedure near zero temperature.

 The physical meaning of Eq. \ref{taueqgene}
is that the equilibration time is the sum of $N$ random variables, possibly
sligthly correlated (the same random coupling $J_{n,n+1}$ appears in the two terms
corresponding to $k=n$ and $k=n+1$). For the Glauber case of Eq. \ref{taueqglauber},
even these slight correlations disappear and the equilibration time
reduces to the sum of independent random variables $e^{2 \beta J_{k,k+1}  } $
whose distribution can be computed from the distribution of the random couplings $J_{k,k+1}$.

Note that in the limit of a pure chain where all ferromagnetic coupling 
have the same value $J$, Eq \ref{taueqgene} reduces to
\begin{eqnarray}
t_{eq}^{pure}(N) 
&& = \frac{1}{ 4 }  
\left[ 2 \frac{e^{ \beta J }}{G \left(J \right)}
+  (N-2)  \frac{e^{ 2 \beta J }}{G \left(0 \right)} 
\right] 
\propto \frac{ N  e^{ 2 \beta J } }{G \left(0 \right)}
\label{taueqgenepure}
\end{eqnarray}
as expected : the Arrhenius factor $e^{ 2 \beta J }$ comes from the barrier $(2J)$
 to create a domain-wall at one boundary, and the prefactor $N$ comes 
from the small probability of order $1/N$ that this domain-wall crosses the whole system of size $N$ instead of returning back.
For the Glauber case where $G(0)=2$, Eq. \ref{taueqgenepure} is in agreement with 
the leading term near zero temperature of the open pure chain
discussed in Refs \cite{pur1d}.

\section{ Boundary Renormalization for the Ising model on the Cayley tree }

\label{sec_tree}

In this section, we consider the pure ferromagnetic Ising model with the classical energy
 of Eq \ref{Uspin} defined on Cayley tree 
of branching ratio $K$ (coordinence $(K+1)$) with $N$ generations, and
with free boundary conditions on all the boundary spins.
We focus on the dynamics corresponding to the simplest choice of Eq. \ref{Wsimple},
where the corresponding quantum Hamiltonian of Eq. \ref{Hsimple} has initial amplitude
$G_k$ all equal to unity. 

We wish to define a closed boundary RG procedure that preserves the symmetry between
the $K$ offsprings of a given site.
So the basic RG step concerns 
$K$ renormalized boundary spins $(S_1,S_2,..,S_K)$
whose renormalized dynamics is described by some renormalized amplitude $G$ 
(which is a number and not an operator) and their common ancestor spin $S$
whose dynamics is still described by the initial amplitude $g=1$ and by
the external field $B_a=J_a \sigma^z_a$ induced by its next ancestor spin $S_a$.
The ferromagnetic couplings 
have still their initial value $J$ so that we have to study the following effective
Hamiltonian for these $(K+1)$ spins $(S_1,..,S_K,S)$

\begin{eqnarray}
H_{K+1}^{simple} \equiv  \left( e^{ - \beta \sigma^z \left( \sum_{i=1}^K J \sigma^z_i+B_a \right)  } -   \sigma^x \right)
+ G \sum_{i=1}^K  \left( e^{ - \beta \sigma^z_i J \sigma^z  } -   \sigma^x_i \right) 
\label{HK}
\end{eqnarray}
After the first RG step, the renormalized amplitude $G$ is expected to become smaller
and smaller, so the most appropriate approach is a perturbative analysis
in the parameter $G$.

\subsection{ Properties of the Hamiltonian $ H_{K+1}^{simple}$ for $G=0$ }

For $G=0$, the spins $(S_1,..,S_K)$ cannot flip and are thus frozen,
so the problem reduces to the single spin $S$ in the external field
 $B=\left(\sum_{i=1}^K J S_i +B_a \right) $ that we have already studied in section 
\ref{elementaryoperator}.
So the $2^{K+1}$ states can be decomposed into

(i) the $2^K$ states corresponding to Eq. \ref{vknormadef}
\begin{eqnarray}
\vert v^{S_1,S_2,..,S_K}_0 > && \equiv \prod_{j=1}^K \vert S_j>  \sum_{S=\pm} 
\frac{e^{ \frac{\beta}{2} S (\sum_{i=1}^K JS_i+B_a) }}{ \sqrt{ 2 \cosh \beta (\sum_{i=1}^K J S_i+B_a)} }  \vert S > 
\label{vkbtree}
\end{eqnarray}
that have a vanishing eigenvalue for $G=0$
\begin{eqnarray}
 \left( e^{ - \beta \sigma^z \left( \sum_{i=1}^K J \sigma^z_i+B_a \right) }-   \sigma^x \right)
\vert v^{S_1,S_2,..,S_K}_0 > && =0
\label{HKtreev}
\end{eqnarray}
The physical interpretation is that the spin $S$ is at equilibrium with respect to the
frozen spins $(S_1,..,S_K)$.

(ii) the $2^K$ states corresponding to Eq. \ref{wknormadef}
\begin{eqnarray}
\vert w^{S_1,S_2,..,S_K}_0 > && \equiv \prod_{j=1}^K \vert S_j> \sum_{S=\pm} 
\frac{ S e^{ - \frac{\beta}{2} S (\sum_{i=1}^K JS_i+B_a) }}{\sqrt{ 2 \cosh \beta (\sum_{i=1}^K JS_i+B_a)} }  \vert S > 
\label{wkbtree}
\end{eqnarray}
that have a finite eigenvalue for $G=0$
\begin{eqnarray}
  \left( e^{ - \beta \sigma^z \left( \sum_{i=1}^K J \sigma^z_i+B_a \right) }-   \sigma^x \right)
\vert w^{S_1,S_2,..,S_K}_0 > && = \left[ 2 \cosh \beta (\sum_{i=1}^K JS_i+B_a) \right] \ \  \vert w^{S_1,S_2,..,S_K}_0 > 
\label{H3treewk}
\end{eqnarray}

Note that the exact ground state (Eq \ref{psi0}) of the Hamiltonian of Eq. \ref{HK}
\begin{eqnarray}
\vert \psi_0> && =\frac{1}{\sqrt{Z_{K+1} }}  \sum_{S_1=\pm} \sum_{S_2=\pm}.. \sum_{S_K=\pm} \sum_{S=\pm}
e^{ \frac{\beta}{2} S (\sum_{i=1}^K JS_i+B_a)}  \vert S_1> \vert S_2>...\vert S_K>  \vert S> 
\nonumber \\ 
Z_{K+1} && \equiv  \sum_{S_1=\pm} \sum_{S_2=\pm} ....\sum_{S_k=\pm}\sum_{S=\pm}
e^{ \beta S (\sum_{i=1}^K JS_i+B_a)}
= \sum_{S_1=\pm} \sum_{S_2=\pm} ... .\sum_{S_k=\pm}\left[ 2 \cosh \beta (\sum_{i=1}^K JS_i+B_a) \right]
\label{psizeroKtree}
\end{eqnarray}
belongs to the subspace spanned by the $2^K$ states $\vert v^{S_1,S_2,..,S_K}_0 > $ (Eq \ref{HKtreev})
\begin{eqnarray}
\vert \psi_0> && =\frac{1}{\sqrt{Z_{K+1} }}  \sum_{S_1=\pm} \sum_{S_2=\pm} ...\sum_{S_K=\pm}
 \sqrt{ 2 \cosh \beta (\sum_{i=1}^K JS_i+B_a)}   \vert v^{S_1,S_2,..,S_K}_0 >
\label{psizeroKtreev}
\end{eqnarray}

\subsection{ Perturbation in the parameter $G$ }

We have seen that for $G=0$, there are $2^K$ states (Eq \ref{HKtreev}) that have zero energy.
For small $G>0$, the perturbation will lift this degeneracy :
the exact ground state of Eq. \ref{psizeroKtreev} will keep its zero energy for arbitrary $G$,
but the other $(2^K-1)$ eigenvalues will become positive as soon as $G>0$.
To determine them, we need to diagonalize the perturbation within the subspace 
spanned by the $2^K$ vectors $\vert v^{S_1,S_2,...S_K}_0 >$, i.e. we look for an eigenstate via the linear combination
\begin{eqnarray}
\vert u_{\lambda} > && = \sum_{S_1=\pm,S_2=\pm,..,S_K=\pm} T^{S_1, S_2,...,S_K} \vert v^{S_1,S_2,...S_K}_0 >
\label{vktree}
\end{eqnarray}
The eigenvalue-equation
\begin{eqnarray}
0 && = (H_{K+1}^{simple}-\lambda) \vert u_{\lambda} > 
\nonumber \\
&& = \sum_{S_1=\pm,S_2=\pm,..,S_K=\pm} T^{S_1, S_2,...,S_K}
\left[ \sum_{i=1}^K G \left( e^{ - \beta \sigma^z_i J \sigma^z  } -   \sigma^x_i \right)   -\lambda \right]
\vert v^{S_1,S_2,...S_K}_0 > 
\label{eigenperturb}
\end{eqnarray}
can be projected onto the $2^K$ bra $< v^{S_1',S_2',..S_K'}_0 \vert$
to obtain a system of $2^K$ linear equations for the $2^K$
 coefficients $T^{S_1, S_2,...,S_K} $
\begin{eqnarray}
0 && = < v^{S_1',..S_K'}_0 \vert (H_{K+1}^{simple}-\lambda) \vert u_{\lambda} > 
\nonumber \\
&& =< v^{S_1',..S_K'}_0 \vert  \sum_{S_1=\pm,...,S_K} T^{S_1, S_2,...,S_K}
\left[ \sum_{i=1}^K G \left( e^{ - \beta \sigma^z_i J \sigma^z  } -   \sigma^x_i \right)   -\lambda \right]
\vert v^{S_1,...S_K}_0 > 
\label{eigenperturbproj}
\end{eqnarray}
From the matrix elements
\begin{eqnarray}
&&  < v^{S_1'S_2',..S_K'}_0 \vert  \left[ e^{ - \beta \sigma^z_i J \sigma^z  } -   \sigma^x_i \right]  
\vert v^{S_1,S_2,...S_K}_0 > 
 = \left( \prod_{j } \delta_{S_j',S_j} \right)
 \frac{ 2 \cosh \beta  (\sum_{j\ne i} JS_j+B_a) }{  2 \cosh \beta (\sum_{j=1}^K J S_j+B_a) } 
\nonumber \\
&& - \left( \prod_{j \ne i} \delta_{S_j',S_j} \right) \delta_{S_i',-S_i}
\frac{ 2 \cosh  \beta   (\sum_{j\ne i} JS_j+B_a) }
{\sqrt{ 2 \cosh \beta (\sum_{j\ne i} JS_j -J S_i
+B_a)}  \sqrt{ 2 \cosh \beta (\sum_{j\ne i} J S_j+JS_i+B_a)} }
\label{matrixi}
\end{eqnarray}
we obtain that Eq \ref{eigenperturbproj} yields
\begin{eqnarray}
0  && =  \sum_{S_1=\pm,...,S_K=\pm} T^{S_1, ...,S_K}
\left[ \sum_{i=1}^K G 
< v^{S_1',..S_K'}_0 \vert \left( e^{ - \beta \sigma^z_i J \sigma^z  } -   \sigma^x_i \right) 
\vert v^{S_1,...S_K}_0 >   -\lambda < v^{S_1',..S_K'}_0 \vert  v^{S_1,...S_K}_0 > \right]
\nonumber \\
&&
= \left[ G \sum_{i=1}^K \frac{ 2 \cosh \beta  (\sum_{j\ne i} JS_j'+B_a) }{  2 \cosh \beta (\sum_{j=1}^K J S_j'+B_a) } -\lambda  \right] T^{S_1',,...,S_K'}
\nonumber \\
&& - G \sum_{i=1}^K \frac{ 2 \cosh  \beta   (\sum_{j\ne i} JS_j'+B_a) }
{\sqrt{ 2 \cosh \beta (\sum_{j\ne i} JS_j' -J S_i'
+B_a)}  \sqrt{ 2 \cosh \beta (\sum_{j\ne i} J S_j'+JS_i'+B_a)} }
T^{S_1', .. ,-S_i',...,S_K'}
\label{eigenperturbprojexpli}
\end{eqnarray}

Let us now use the symmetry between the $K$ branches
to note $t(k)$ the components $T^{S_1', S_2',...,S_K'} $ where $k$ spins take the value $(+)$
and $(K-k)$ take the value $-$, i.e.
\begin{eqnarray}
 T^{S_1', S_2',...,S_K'} = t \left(k=\sum_{i=1}^K \frac{1+S_i'}{2} \right)
\label{rksym}
\end{eqnarray}
For the two extremal cases  $t(K)$ (all spins are $(+)$) and $t(0)$ (all spins are $(-)$),
Eq \ref{eigenperturbprojexpli} becomes
\begin{eqnarray}
0  &&   
= \left[ G K \frac{ 2 \cosh \beta  \left(  (K-1)J+B_a \right) }
{  2 \cosh \beta (K J +B_a) } -\lambda  \right] t(K)
 - G K \frac{ 2 \cosh  \beta    \left( (K-1) J+B_a \right) }
{\sqrt{ 2 \cosh \beta (  \left(K-2)J
+B_a \right)}  \sqrt{ 2 \cosh \beta (KJ+B_a)} }
t(K-1)
\nonumber \\
0  &&   
= \left[ G K \frac{ 2 \cosh \beta  \left(  (K-1)J-B_a \right) }
{  2 \cosh \beta (K J -B_a) } -\lambda  \right] t(0)
 - G K \frac{ 2 \cosh  \beta    \left( (K-1) J-B_a \right) }
{\sqrt{ 2 \cosh \beta (  \left(K-2)J
-B_a \right)}  \sqrt{ 2 \cosh \beta (KJ-B_a)} }
t(1)
\label{eigensymferro}
\end{eqnarray}
whereas for the non-extremal case $0<k<K$, Eq. \ref{eigenperturbprojexpli} reads
\begin{eqnarray}
0  
&&
= \left[ G k  \frac{ 2 \cosh \beta  ( (2k-K-1)J+B_a) }{  2 \cosh \beta ( (2k-K)J+B_a) }
+ G (K-k) \frac{ 2 \cosh \beta  ( (2k-K+1)J+B_a) }{  2 \cosh \beta ( (2k-K)J+B_a) }
 -\lambda  \right] t(k)
\nonumber \\
&& - G k \frac{ 2 \cosh  \beta   ((2k-K-1)J +B_a) }
{\sqrt{ 2 \cosh \beta ( (2k-K-2)J +B_a)}  \sqrt{ 2 \cosh \beta ( (2k-K)J+B_a)} }
t(k-1)
\nonumber \\
&& - G (K-k) \frac{ 2 \cosh  \beta   ( (2k-K+1)J +B_a) }
{\sqrt{ 2 \cosh \beta ( (2k-K+2)J+B_a)}  \sqrt{ 2 \cosh \beta ( (2k-K)J +B_a)} }
 t(k+1)
\label{eigensyminter}
\end{eqnarray}
It is easy to check that the components (not normalized here) of the exact 
ground state of Eq \ref{psizeroKtree}
\begin{eqnarray}
t_0(k)  && = \sqrt{ 2 \cosh \beta ( (2k-K) J+B_a)}  
\label{psizero3treecompo}
\end{eqnarray}
satisfy Equations \ref{eigensymferro} and \ref{eigensyminter} for $\lambda=0$ as it should.

To compute the lowest non-vanishing eigenvalue $\lambda_1$,
it is consistent to set $\lambda_1=0$ in all eqs \ref{eigensyminter}
concerning the non-extremal cases $0<k<K$, leading to
\begin{eqnarray}
0  
&&
= \left[  k  \frac{ 2 \cosh \beta  ( (2k-K-1)+B_a) }{ \sqrt{ 2 \cosh \beta ( (2k-K)J+B_a)} }
+  (K-k) \frac{ 2 \cosh \beta  ( (2k-K+1)+B_a) }{ \sqrt{ 2 \cosh \beta ( (2k-K)J+B_a)} }
  \right] t(k)
\nonumber \\
&& -  k \frac{ 2 \cosh  \beta   ((2k-K-1) +B_a) }
{\sqrt{ 2 \cosh \beta ( (2k-K-2) +B_a)}   }
t(k-1)
 -  (K-k) \frac{ 2 \cosh  \beta   ( (2k-K+1) +B_a) }
{\sqrt{ 2 \cosh \beta ( (2k-K+2)+B_a)}  }
 t(k+1)
\label{eigensyminterlambda1zero}
\end{eqnarray}
that may be solved in terms of the boundary conditions $t(K)$ and $t(0)$.

To obtain the explicit solution, it is convenient to introduce the amplitudes $A(k)$
with respect to the ground state components of Eq. \ref{psizero3treecompo}
\begin{eqnarray}
t(k) \equiv A(k) t_0(k)  && = A(k) \sqrt{ 2 \cosh \beta ( (2k-K) J+B_a)}  
\label{defamplirA}
\end{eqnarray}
so that Eq. \ref{eigensyminterlambda1zero} takes the simpler form
\begin{eqnarray}
  A(k)   = p_-(k)  A(k-1) +p_+(k) A(k+1)  
\label{eigensymampli}
\end{eqnarray}
with the notations
\begin{eqnarray}
p_-(k)   && \equiv \frac{ k  2 \cosh  \beta   ((2k-K-1) +B_a)}{\left[  k   2 \cosh \beta  ( (2k-K-1)+B_a) 
+  (K-k)  2 \cosh \beta  ( (2k-K+1)+B_a) 
  \right]}
\nonumber \\
p_+(k)   && \equiv \frac{(K-k)  2 \cosh  \beta   ( (2k-K+1) +B_a)}
{\left[  k   2 \cosh \beta  ( (2k-K-1)+B_a) 
+  (K-k)  2 \cosh \beta  ( (2k-K+1)+B_a) 
  \right]} = 1-p_-(k)
\label{eigensymamplippm}
\end{eqnarray}
Let us introduce two linearly independent solutions.
The solution corresponding to the boundary conditions
\begin{eqnarray}
  Q_K(0)  && = 0 
\nonumber \\
 Q_K(K)  && = 1 
\label{bcsolQK}
\end{eqnarray}
 can be obtained by recurrence \cite{rec1d} and reads 
\begin{eqnarray}
  Q_K(k)  && = \frac{R_K(1,k)}{R_K(1,K)}
\label{solQK}
\end{eqnarray}
using Kesten variables \cite{kestenv} 
\begin{eqnarray}
R_K(1,0) && =0 \nonumber \\
R_K(1,1) && =1 \nonumber \\
R_K(1,k \geq 2) && =1+\sum_{m=1}^{k-1} \prod_{n=1}^{m} \frac{ p_-(n)}{ p_+(n)}
\nonumber \\
R_K(1,K) && = 1+\sum_{m=1}^{K-1} \prod_{n=1}^{m} \frac{ p_-(n)}{ p_+(n)}
 = 1+ \frac{p_-(1)}{p_+(1)} + \frac{p_-(1)p_-(2)}{p_+(1)p_+(2)} + ... + \frac{p_-(1) p_-(2)...p_-(K-1)}{p_+(1) p_+(2)...p_+(K-1)}
\label{rsoluexitchaint}
\end{eqnarray}
Similarly, the solution corresponding to the boundary conditions
\begin{eqnarray}
  Q_0(0)  && = 1
\nonumber \\
 Q_0(K)  && = 0 
\label{bcsolQKzero}
\end{eqnarray}
reads 
\begin{eqnarray}
  Q_0(k)  && = \frac{R_0(k,K-1)}{R_0(0,K-1)}
\label{solQKzero}
\end{eqnarray}
with
\begin{eqnarray}
R_0(K,K-1) && =0  \\
R_0(K-1,K-1) && =1 \nonumber \\
R_0(k \leq K-2 , K-1) && =1+\sum_{m=k+1}^{K-1} \prod_{n=m}^{K-1} \frac{ p_+(n)}{ p_-(n)}
\nonumber \\
R_0(0,K-1) && = 1+\sum_{m=1}^{K-1} \prod_{n=m}^{K-1} \frac{ p_+(n)}{ p_-(n)}
 = 1+ \frac{p_+(K-1)}{p_-(K-1)} 
+ ... +
 \frac{p_+(K-1) p_+(K-2)...p_+(1)}{p_-(K-1) p_-(K-2)...p_-(1)}
\nonumber 
\label{rsoluexitchaintbis}
\end{eqnarray}

It is useful to introduce the continuation of the ground state components of
Eq \ref{psizero3treecompo} to half-integers to rewrite 
 the ratios
\begin{eqnarray}
\frac{ p_-(k) }{p_+(k)}   && = \frac{ k  2 \cosh  \beta   ((2k-K-1) +B_a)}
{  (K-k)  2 \cosh \beta  ( (2k-K+1)+B_a) }
= \frac{ k t_0^2 \left( k-\frac{1}{2} \right) }{(K-k) t_0^2 \left( k+\frac{1}{2} \right)}
\label{ratioppm}
\end{eqnarray}
and the products 
\begin{eqnarray}
\prod_{n=1}^{m} \frac{ p_-(n)}{ p_+(n)} && = 
\prod_{n=1}^{m} \left[  \frac{ n t_0^2 \left( n-\frac{1}{2} \right) }{(K-n) t_0^2 \left( n+\frac{1}{2} \right)} \right]
= \frac{ m! (K-1-m)! }{(K-1)! } \ \frac{t_0^2 \left( \frac{1}{2} \right)}{t_0^2 \left( m+\frac{1}{2} \right)}
\nonumber \\
 \prod_{n=m}^{K-1} \frac{ p_+(n)}{ p_-(n)}
&& = \prod_{n=m}^{K-1}  \frac{(K-n) t_0^2 \left( n+\frac{1}{2} \right)}{ n t_0^2 \left( n-\frac{1}{2} \right) }
= \frac{ (m-1)! (K-m)! }{(K-1)! } \ \frac{t_0^2 \left( K-\frac{1}{2} \right)}
{t_0^2 \left( m-\frac{1}{2} \right)}
\label{productsratioppm}
\end{eqnarray}

In particular in the following, we will need the two denominators
\begin{eqnarray}
R_K(1,K) && = 1+\sum_{m=1}^{K-1} \frac{ m! (K-1-m)! }{(K-1)! } \ \frac{t_0^2 \left( \frac{1}{2} \right)}{t_0^2 \left( m+\frac{1}{2} \right)}
 =t_0^2 \left( \frac{1}{2} \right)
\sum_{m=0}^{K-1}  \ \frac{1}{C^{m}_{K-1} t_0^2 \left( m+\frac{1}{2} \right)}
\label{deno1}
\end{eqnarray}
and
\begin{eqnarray}
R_0(0,K-1) && 
 =  1+\sum_{m=1}^{K-1}  \frac{ (m-1)! (K-m)! }{(K-1)! } \ \frac{t_0^2 \left( K-\frac{1}{2} \right)}
{t_0^2 \left( m-\frac{1}{2} \right)}
=t_0^2 \left( K-\frac{1}{2} \right) \sum_{m=0}^{K-1}  
 \frac{1}
{ C^{m}_{K-1} t_0^2 \left( m+\frac{1}{2} \right)}
\label{deno2}
\end{eqnarray}
that determine the solutions near the boundaries of $Q_K$
\begin{eqnarray}
  Q_K(1)  && = \frac{R_K(1,1)}{R_K(1,K)}=\frac{1}{R_K(1,K)}
\nonumber \\
1-Q_K(K-1) && = \frac{R_K(1,K) - R_K(1,K-1)}{R_K(1,K)} = \frac{1}{R_0(0,K-1)}
\label{solQKspecial}
\end{eqnarray}
and of $Q_0$
\begin{eqnarray}
  Q_0(K-1)  && = \frac{R_0(K-1,K-1)}{R_0(0,K-1)} = \frac{1}{R_0(0,K-1)}
\nonumber \\
1-Q_0(1) && = \frac{R_0(0,K-1)- R_0(1,K-1)}{R_0(0,K-1)} = \frac{1}{R_K(1,K)}
\label{solQKspecial2}
\end{eqnarray}

The solution of the system \ref{eigensymampli} 
that satisfy the boundary conditions (Eq \ref{defamplirA})
\begin{eqnarray}
 A(0) = \frac{t(0)}{t_0(0)}  
 \nonumber \\
 A(K) = \frac{t(K)}{t_0(K)}  
\label{bcA}
\end{eqnarray}
can be obtained by the linear combination
\begin{eqnarray}
  A(k)  && = A(0)  Q_0(k) +  A(K)  Q_K(k) =  \frac{t(0)}{t_0(0)} Q_0(k)+\frac{t(K)}{t_0(K)}
Q_K(k)
\label{solAKzero}
\end{eqnarray}
so that the solution of the system \ref{eigensyminterlambda1zero} 
reads 
\begin{eqnarray}
t(k) = A(k) t_0(k)  && = \left[  \frac{t(0)}{t_0(0)} Q_0(k)+\frac{t(K)}{t_0(K)}
Q_K(k) \right] t_0(k)
\label{solurk}
\end{eqnarray}

To determine $\lambda_1$, we just need to replace 
\begin{eqnarray}
t(1)
&&= \left[  \frac{t_0(1)}{t_0(0)} Q_0(1) \right] t(0)
+  \left[  \frac{t_0(1)}{t_0(K)} Q_K(1) \right] t(K)
\nonumber \\
t(K-1)  
&& =\left[  \frac{t_0(K-1)}{t_0(0)} Q_0(K-1) \right]  t(0)
+ \left[ \frac{t_0(K-1)}{t_0(K)} Q_K(K-1) \right]  t(K)
\label{solurk1}
\end{eqnarray}
in Eqs \ref{eigensymferro} to obtain 
the following system of two linear equations
for the two components $t_{\lambda_1}(0)$ and $t_{\lambda_1}(K)$
\begin{eqnarray}
0  &&   
 = \left[ 1-  Q_K(K-1) -\lambda_1 \frac{ t_0^2(K)}{G K  t_0^2 \left( K-\frac{1}{2} \right)}  \right] t_{\lambda_1}(K)
 -  \frac{ t_0(K)}{t_0(0)} Q_0(K-1)   t_{\lambda_1}(0)
\nonumber \\
0  
&& = \left[ 1 - Q_0(1)-\lambda_1 \frac{ t_0^2(0) }{G K t_0^2  \left( \frac{1}{2}  \right) }
  \right] t_{\lambda_1}(0)
 -  \frac{ t_0(0) }{t_0(K)} Q_K(1)  t_{\lambda_1}(K)
\label{eigensymferrotwoeff}
\end{eqnarray}

The two components (not normalized here) are orthogonal to Eq \ref{psizero3treecompo}
as it should and read
\begin{eqnarray}
t_{\lambda_1}(0) && = t_0(K)   = \sqrt{ 2 \cosh \beta ( K J+B_a)} 
\nonumber \\
 t_{\lambda_1}(K) && = - t_0(0)   = - \sqrt{ 2 \cosh \beta ( KJ-B_a)} 
\label{psiun3treecompo}
\end{eqnarray}
The corresponding eigenvalue $\lambda_1$ reads
 using Eqs \ref{deno1}, \ref{deno2}, \ref{solQKspecial}, Eq \ref{solQKspecial2}
\begin{eqnarray}
\lambda_1 
&& =  \frac{G K  } {
\sum_{m=0}^{K-1}  \ \frac{1}{C^{m}_{K-1}
2 \cosh \beta ( (2m+1-K) J+B_a)   }   }
 \left[  \frac{1}{2 \cosh \beta ( K J-B_a) }
 +  \frac{ 1 }{ 2 \cosh \beta ( K J+B_a) }  \right]
\label{lambda1prefinal}
\end{eqnarray}
Since this expression is unchanged via the transformation $B_a=J S_a \to -B_a$,
we may replace $B_a$ by its absolute value $\vert B_a \vert=J$ to obtain the final expression
for the lowest non-vanishing eigenvalue $\lambda_1$ at first order in perturbation
with respect to the parameter $G$
\begin{eqnarray}
\lambda_1 
&& =  \frac{G K  } {
\sum_{m=0}^{K-1}  \ \frac{1}{C^{m}_{K-1}
2 \cosh \beta J  (2m+2-K)    }   }
 \left[  \frac{1}{2 \cosh \beta J  (K-1) }
 +  \frac{ 1 }{ 2 \cosh \beta J  (K+1)  }  \right] +O(G^2)
\label{lambda1final}
\end{eqnarray}

\subsection{ Renormalization rule for the amplitude $G^R$ }

Let us now project the Hamiltonian of Eq. \ref{HK}
onto its two lowest eigenvalues $\lambda_0=0$ and $\lambda_1$
\begin{eqnarray}
H^{simple}_{K+1} \simeq \frac{\lambda_1}{ < u_{\lambda_1}\vert u_{\lambda_1} >}
  \vert u_{\lambda_1} > < u_{\lambda_1}\vert 
\label{Hefflowest}
\end{eqnarray}
where the eigenvector of Eq. \ref{vktree}
can be approximated at low temperature by its two components onto
fully ferromagnetic states
\begin{eqnarray}
  \vert u_{\lambda_1} > && \simeq  t_{\lambda_1}(K) \left(  \prod_{j=1}^K \vert S_j=1> \right) \sum_{S=\pm} 
\frac{e^{ \frac{\beta}{2} S  (KJ +B_a) }}{ \sqrt{ 2 \cosh \beta (KJ +B_a)} }  \vert S > 
 \nonumber \\ && 
+ t_{\lambda_1}(0) \left(  \prod_{j=1}^K \vert S_j=-1> \right)
\sum_{S=\pm} 
\frac{e^{ \frac{\beta}{2} S  (-KJ +J_a) }}{ \sqrt{ 2 \cosh \beta (KJ -B_a)} }  \vert S > 
\nonumber \\
&& \simeq   t_{\lambda_1}(K) \left(  \prod_{j=1}^K \vert S_j=1> \right)   \vert S=+1 > 
+ t_{\lambda_1}(0) \left(  \prod_{j=1}^K \vert S_j=-1> \right)  \vert S=-1 > 
\label{ulambda1}
\end{eqnarray}
with the coefficients (Eq \ref{psiun3treecompo})
\begin{eqnarray}
t_{\lambda_1}(0) && = \sqrt{ 2 \cosh \beta ( K J+B_a)} \simeq e^{ \frac{\beta}{2} (KJ+ B_a)}
\nonumber \\
 t_{\lambda_1}(K) && = -  \sqrt{ 2 \cosh \beta ( K J-B_a)}  \simeq -  e^{ \frac{\beta}{2} (KJ- B_a)}
\label{psizero3treecompolowT}
\end{eqnarray}
Finally at leading order near zero temperature, one obtains
\begin{eqnarray}
  \vert u_{\lambda_1} > && \simeq e^{ \frac{\beta}{2} KJ }
\left[  e^{ \frac{\beta}{2} B_a} \left(  \prod_{j=1}^K \vert S_j=1> \right)   \vert S=+1 > 
-  e^{- \frac{\beta}{2}  B_a} \left(  \prod_{j=1}^K \vert S_j=-1> \right)  \vert S=-1 > \right]
\label{ulambda1lowT}
\end{eqnarray}
with the corresponding normalization (using $B_a=J S_a$)
\begin{eqnarray}
 < u_{\lambda_1}\vert u_{\lambda_1} > 
&& \simeq  e^{ \beta (KJ+ B_a)} + e^{ \beta (KJ - B_a)} = e^{ \beta (KJ)} 2 \cosh (\beta B_a)
=  e^{ \beta (KJ)} 2 \cosh (\beta J)
\label{normauu}
\end{eqnarray}

In terms of the renormalized spin
\begin{eqnarray}
\vert S_R=+> && \equiv  \left(  \prod_{j=1}^K \vert S_j=1> \right)   \vert S= + > \nonumber \\
\vert S_R=-> && \equiv  \left(  \prod_{j=1}^K \vert S_j=-1> \right)   \vert S= - > 
\label{SR}
\end{eqnarray}
and of the external local field $B_R=B_a=J S_a$, the effective Hamiltonian of Eq. \ref{Hefflowest}
can be rewritten as an elementary operator (Eq. \ref{Hsimplesumlocalproj})
\begin{eqnarray}
H^{simple}_{K+1} \simeq G_R  \left(e^{  -  \frac{\beta}{2} B_R } \vert S_R=+>
- e^{   \frac{\beta}{2} B_R }  \vert S_R=->  \right)
 \left(e^{  -  \frac{\beta}{2} B_R } <  S_R=+ \vert
- e^{   \frac{\beta}{2} B_R } <  S_R=- \vert   \right)
\label{Hefflowestfinal}
\end{eqnarray}
with the renormalized amplitude (using Eq \ref{lambda1final})
 \begin{eqnarray}
G_R &&  =\frac{ \lambda_1 }
{ 2 \cosh    \beta  J   } =
 \frac{G K  } { 2 \cosh    \beta  J
\sum_{m=0}^{K-1}  \ \frac{1}{C^{m}_{K-1}
2 \cosh \beta J  (2m+2-K)    }   }
 \left[  \frac{1}{2 \cosh \beta J  (K-1) }
 +  \frac{ 1 }{ 2 \cosh \beta J  (K+1)  }  \right]
\label{ggenetree}
\end{eqnarray}

To be consistent with the previous low-temperature approximations, 
we now should evaluate the leading behavior of Eq. \ref{ggenetree} near zero temperature,
i.e. we should replace hyperbolic functions by exponentials.
In particular, one has
 \begin{eqnarray}
\frac{1}{ 2 \cosh \beta J  (2m+2-K)    }    = 
\frac{1}{ e^{ \beta J  (2m+2-K) }+ e^{- \beta J  (2m+2-K) }   }   
&& \simeq \frac{1}{2} \ \ {\rm if } \ \ m=\frac{K}{2} -1
\nonumber \\
&& \simeq e^{ - \beta J \vert 2m+2-K \vert} \ {\rm if } \ \ m \ne \frac{K}{2}-1 
\label{termchmlowT}
\end{eqnarray}
so that the leading term near low temperature of Eq. \ref{ggenetree}
depends on the parity of $K$.

\subsubsection { Leading behavior near zero temperature for even $K$ }

When the branching ratio $K$ is even, then $\left(\frac{K}{2}-1\right)$ is an integer, so that 
the integer $m$ can take this value, and the sum in the denominator of Eq. \ref{ggenetree}
is dominated by this contribution
 \begin{eqnarray}
\sum_{m=0}^{K-1}  \ \frac{1}{C^{m}_{K-1}
2 \cosh \beta J  (2m+2-K)    }  &&
  \simeq   \ \frac{1}{ 2 C^{\frac{K}{2}-1}_{K-1}    }   
\label{kevensumlowT}
\end{eqnarray}
so that Eq \ref{ggenetree} reads at leading order
 \begin{eqnarray}
{\rm K \ \ even \ :  \ \ \ } G^R  && \simeq  
 G  e^{- \beta J K}   2 K  C^{\frac{K}{2}-1}_{K-1} 
= G  e^{- \beta J K}   2 \frac{ K! }{ \left( \frac{K}{2} \right)!\left( \frac{K}{2}-1 \right)! } 
\label{ggenelowTKeven}
\end{eqnarray}

For instance for $K=2$, one obtains
 \begin{eqnarray}
 K=2  \ :  \ \ \  G^R  && \simeq   4  e^{- 2 \beta J } G
\label{ggenelowTK2}
\end{eqnarray}

\subsubsection { Leading behavior near zero temperature for odd $K$ }

When the branching ratio $K$ is odd, then $\left(\frac{K}{2}-1\right)$ is not an integer
so that $m$ cannot take
this value, and the sum in the denominator is dominated by the contributions
of the two closest integers $m=\frac{K-3}{2}$ and $m=\frac{K-1}{2} $
 \begin{eqnarray}
\sum_{m=0}^{K-1}  \ \frac{1}{C^{m}_{K-1}
2 \cosh \beta J  (2m+2-K)    }  &&
  \simeq \sum_{m=0}^{K-1}
  \ \frac{e^{ - \beta J \vert 2m+2-K \vert}}{C^{m}_{K-1}    }   
\simeq \frac{e^{ - \beta J }}{C^{\frac{K-3}{2}}_{K-1}    } 
+ \frac{e^{ - \beta J }}{C^{\frac{K-1}{2}}_{K-1}    } 
 \nonumber \\ &&  = 
e^{ - \beta J } \left[ 
\frac{ \left(\frac{K+1}{2} \right)! \left(\frac{K-3}{2} \right)!  }{ (K-1)!}
+ \frac{\left(\frac{K-1}{2} \right)! \left(\frac{K-1}{2} \right)!  }{ (K-1)!}  \right]
 \nonumber \\ &&  = 
e^{ - \beta J } 
\frac{ K \left(\frac{K-1}{2} \right)! \left(\frac{K-3}{2} \right)!  }{ (K-1)!}
\label{koddsumlowT}
\end{eqnarray}

so that Eq \ref{ggenetree} reads at leading order
 \begin{eqnarray}
{\rm K \ \ odd \ :  \ \ \ } G^R  &&
 \simeq G  e^{- \beta J (K-1) }    
\frac{ (K-1)!}{  \left(\frac{K-1}{2} \right)! \left(\frac{K-3}{2} \right)!  }
= G  e^{- \beta J (K-1) }   (K-1) C^{\frac{K-1}{2}}_{K-2}
\label{ggenelowTKodd}
\end{eqnarray}

For instance for $K=3$, one obtains
 \begin{eqnarray}
 K=3  \ :  \ \ \  G^R  && \simeq  
 2  e^{- 2 \beta J  } G
\label{ggenelowTK3}
\end{eqnarray}

\subsection{ Conclusion for the equilibrium time $t_{eq}^{simple}(N)$ of a Cayley tree with $N$ generations }

Let us now consider a finite Cayley tree of branching ratio $K$ with $N$ generations.
For the first RG step where $G_0=1$, we cannot use the perturbative analysis presented above
to obtain $G_1$. However since $G_1 \ll 1$ at low temperature, we may use 
the perturbative analysis given above to obtain the recursion
 \begin{eqnarray}
G_{n+1} \simeq  \frac{ G_{n} }{\rho(K)}  \simeq \frac{ G_1 }{ \left[ \rho(K)\right]^n }
\label{recGtree}
\end{eqnarray}
for all RG steps corresponding $1 \leq n \leq N-1$, where the factor $\rho(K) $
has been evaluated at low temperature (Eqs \ref{ggenelowTKeven} and \ref{ggenelowTKodd})
 \begin{eqnarray}
\rho(K) &&  =    \frac{e^{ \beta J K}}{   2 K  C^{\frac{K}{2}-1}_{K-1} }
= e^{ \beta J K} \frac{ \left( \frac{K}{2} \right)!\left( \frac{K}{2}-1 \right)! } { 2( K!) }
 \ \ \  \ \ \ \ \ \ \ \ {\rm for \ even } \ \ K 
 \nonumber \\ &&
 =   \frac{ e^{ \beta J (K-1) } }{ (K-1)  C^{\frac{K-1}{2}}_{K-2} }
=  e^{ \beta J (K-1) } \frac{  \left(\frac{K-1}{2} \right)! \left(\frac{K-3}{2} \right)!  }{ (K-1)!}
 \ \ \ \ \ {\rm for \ odd } \ \ K 
\label{rhoK}
\end{eqnarray}
Finally, at the last RG step, we could take into account that the center has $(K+1)$ neighbors instead
of $K$, and has no further ancestor $B_a=0$. However, this anomalous last step is only a boundary 
multiplicative contribution, as is $G_1$, and cannot change the dependence upon the number $N$ of generations
for large $N$ coming from Eq \ref{recGtree}
 \begin{eqnarray}
G_{N}^{final} \propto \frac{1}{\left[ \rho(K) \right]^N}
\label{Gfinaltree}
\end{eqnarray}

In summary, we obtain that the equilibrium time $t_{eq}^{simple}(N)=1/(2 G_{N}^{final} )$ (Eq \ref{taueqGfinal})
of a Cayley tree of branching ratio $K$
grows exponentially with the number $N$ of generations
 \begin{eqnarray}
t_{eq}^{simple}(N) \propto \left[ \rho(K) \right]^N
\label{teqtree}
\end{eqnarray}
where the growth factor $\rho(K)$ is given explicitly by Eq \ref{rhoK} for any $K$.

\subsection{  Equilibrium time $t_{eq}^{Glauber}(N)$ for the Glauber dynamics}

The above results concerning the simple dynamics can be extended to the Glauber dynamics as follows.
The Hamiltonian $H_{K+1}^{simple}$ of Eq. \ref{HK} has to be replaced for the first step by
\begin{eqnarray}
H_{K+1}^{Glauber} \equiv \frac{1}{2 \cosh (\beta \left( \sum_{i=1}^K J \sigma^z_i+B_a \right)) } \left( e^{ - \beta \sigma^z \left( \sum_{i=1}^K J \sigma^z_i+B_a \right)  } -   \sigma^x \right)
+ \frac{1}{2 \cosh (\beta J \sigma^z) } \sum_{i=1}^K  \left( e^{ - \beta \sigma^z_i J \sigma^z  } -   \sigma^x_i \right) 
\label{HKglauber}
\end{eqnarray}
Since the $K$ leaves have no external field and are just linked to $\sigma^z$, the amplitude $\frac{1}{2 \cosh (\beta J \sigma^z) } $ reduces to the number $ \frac{1}{2 \cosh (\beta J ) }$. The remaining non-trivial amplitude $\frac{1}{2 \cosh (\beta \left( \sum_{i=1}^K J \sigma^z_i+B_a \right)) } $ 
will disappear when we apply the perturbation method within the subspace annihilating the corresponding operator
$\left( e^{ - \beta \sigma^z \left( \sum_{i=1}^K J \sigma^z_i+B_a \right)  } -   \sigma^x \right) $.
Our conclusion is thus that the equilibrium time $t_{eq}^{Glauber}(N) $ for the Glauber dynamics
will have exactly the same leading exponential behavior in $N$ as the result of Eq. \ref{teqtree} derived for the simple dynamics
 \begin{eqnarray}
t_{eq}^{Glauber}(N) \propto \left[ \rho(K) \right]^N
\label{teqtreeglauber}
\end{eqnarray}
even if the prefactor can differ (see the discussion on the differences between the equilibrium times of the simple and Glauber dynamics
in Appendix \ref{sec_diff}).

\subsection{ Comparison with previous results on dynamical barriers  }

From Eq. \ref{teqtree} and Eq. \ref{teqtreeglauber}, we obtain that
 the energetic barrier $B_K(N)$ defined as the coefficient of $\beta$ in $\ln t_{eq}(N)$ 
 \begin{eqnarray}
B_K(N) = \lim_{\beta \to +\infty}   \frac{ \ln t_{eq}^{simple}(N) }{ \beta } 
&& = N J K +O(1)\ \ \  \ \ \ \ \ \ \ \ \ \ {\rm for \ even } \ \ K 
 \nonumber \\ 
&& = N J (K-1) +O(1) \ \ \  \ \ \ \ \ {\rm for \ odd } \ \ K 
\label{barriertree}
\end{eqnarray}
grows linearly with the number $N$ of generations (i.e. logarithmically with the number of sites
${\cal N}_N \propto K^N$) in agreement with previous works of physicists \cite{henley,melin,montanari}
and of mathematicians \cite{leng,yanna,marti,berger}.
Besides this correct scaling with $N$, it appears that the slope $(K-1) J $ for odd $K$ of Eq. \ref{barriertree}
coincides with the slope obtained in \cite{melin}, where a so-called 'disjoint strategy' is optimal,
whereas the slope $K J $ for even $K$ of Eq. \ref{barriertree} differs from the slope $J(K-1)$ obtained in \cite{henley,melin},
where a so-called 'non-disjoint strategy' is optimal. We refer to Refs \cite{henley,melin,montanari,leng,yanna} for more explanations
on the differences between disjoint/non-disjoint strategies. For the present work, it is clear that the renormalization procedure
making coherent clusters of spins within sub-trees corresponds to the disjoint strategy.

Finally, besides the Arrhenius factor involving the energetic barrier of Eq \ref{barriertree},
 the present renormalization procedure
predicts explicit combinatorial prefactors for the exponential growth factor $\rho(K)$
(Eq. \ref{rhoK})
that have not been previously discussed in the literature, to the best of our knowledge.

\section{ Conclusion }

\label{sec_conclusion}

In this paper, we have introduced a real-space RG procedure valid near zero-temperature
to evaluate the largest relaxation time of classical random ferromagnets.
We have used the standard mapping between 
the master equations satisfying detailed balanced and quantum Hamiltonians
having an exact zero-energy ground state. 
 The largest relaxation time $t_{eq}$ governing the convergence of the dynamics 
towards the Boltzmann
equilibrium is determined by the lowest non-vanishing eigenvalue $E_1=1/t_{eq}$ of the
quantum Hamiltonian $H$. We have thus defined appropriate real-space RG rules 
for the quantum Hamiltonian to evaluate $E_1$ for finite systems.
 We have described how  
 the renormalization flow can be explicitly solved for the two following cases.

(i) For the one-dimensional random ferromagnetic chain with free boundary conditions,
the largest relaxation time $t_{eq}$ can be expressed in terms of the set of random couplings for various choices of the dynamical transition rates. The validity of these RG results in $d=1$ have been checked by comparison with another approach in Appendix.

(ii) For the pure Ising model on a Cayley tree of branching ratio $K$ (coordinence $(K+1)$),
 we have computed the exponential growth of $t_{eq}(N)$ with the number $N$ of generations.

In a companion paper \cite{us_dyndyson}, we explain how the renormalization flow
can be also explicitly solved for the Dyson hierarchical Dyson Ising model.
In the future, we hope to obtain numerical results for the RG flow in finite dimensions $d>1$.

\section*{ Acknowledgments }

It is a pleasure to thank G. Semerjian for his comments on our work
and for pointing out the references \cite{montanari,leng,yanna,marti,berger}.

\appendix

\section{ Check of the validity of the RG procedure in $d=1$}

\label{sec_app}

In this Appendix, we present another approach to check the results of the RG procedure
obtained in section \ref{sec_brg1d} for the random ferromagnetic chain 

\subsection{ Ansatz for the first excited quantum state in terms of exit probabilities }

\label{sec_ansatz}

\subsubsection{ Eigenequation for $\psi_1$ }

For the quantum Hamiltonian ${\cal H}$ corresponding to the spin-flip dynamics of classical spin models
with the energy of Eq. \ref{Uspin}, the exact ground state of Eq. \ref{psi0}
\begin{eqnarray}
\psi_0 ({\cal C}) = \frac{ e^{- \frac{\beta}{2} U({\cal C}) }}{\sqrt Z} = 
\frac{ e^{- \frac{\beta}{2} \sum_{i<j} J_{ij} S_i S_j }}{\sqrt Z}
\label{psi0spins}
\end{eqnarray}
is invariant under a global flip of all the spins
\begin{eqnarray}
 \psi_0(-C)=\psi_0(C)
\label{psizerosym}
\end{eqnarray}
On the contrary, the first excited state will be antisymmetric under a global flip of all the spins
\begin{eqnarray}
 \psi_1(-C)=-\psi_1(C)
\label{psiunantisym}
\end{eqnarray}
but its modulus is expected to coincide nearly with $\psi_0(C) $ in the two valleys around the two classical ferromagnetic ground states.
It is thus convenient to set
\begin{eqnarray}
\psi_1(C) = \psi_0(C) A(C)
\label{defamplitude}
\end{eqnarray}
and to look for the antisymmetric amplitude $A(C)$ (antisymmetric  under a global flip of all the spins)
\begin{eqnarray}
A(-C)= A(C)
\label{antisymA}
\end{eqnarray}
The eigenvalue equation for the quantum Hamiltonian of Eq. \ref{tight},\ref{eps}, \ref{hopping}
\begin{eqnarray}
{\cal H} \vert \psi_1 > = E_1  \vert \psi_1 >
\label{eigenpsi1e1}
\end{eqnarray}
becomes via the change of variables of Eq. \ref{defamplitude}
\begin{eqnarray}
 \left[W_{out}(C)-E_1 \right] A(C) = \sum_{C'} W(C \to C') A(C')
\label{eqA}
\end{eqnarray}

For a large system where $E_1$ is small, we expect that $E_1 $ can be neglected
 with respect to $W_{out}(C) $ for all configurations different from the two classical ground states,
so that one obtains the approximate equation
\begin{eqnarray}
 A(C) \opsimeq_{E_1 \to 0} \sum_{C'} \frac{W(C \to C')}{W_{out}(C)} A(C')
= \sum_{C'} \pi_C(C') A(C')
\label{eqAe1zero}
\end{eqnarray}
where
\begin{eqnarray}
 \pi_C(C') \equiv \frac{W(C \to C')}{W_{out}(C)} = \frac{W(C \to C')}{\sum_{C''}W(C \to C'')}
\label{pi}
\end{eqnarray}
represents the probability that the first exit from configuration $C$ leads to $C'$
for the master equation of Eq. \ref{master},
with the normalization
\begin{eqnarray}
\sum_{C'} \pi_C(C') =1
\label{normapi}
\end{eqnarray}

\subsubsection{ Relation with exit probabilities }

Exit probabilities are known to satisfy backward master equation similar to Eq. \ref{eqAe1zero} (see for instance the textbooks \cite{gardiner,vankampen,risken}). 
More precisely, in a ferromagnet, one may introduce 
the probability $Q_+(C)$ that the dynamics starting in configuration $C$ reaches 
first the configuration $C_+$ (all spins plus) than the configuration $C_-$
(all spins minus).
The complementary probability $Q_-(C)=1-Q_+(C)$ represents 
the probability that the dynamics starting in configuration $C$ reaches 
first the configuration $C_-$ than the configuration $C_+$.
The escape probability satisfies the backward master equation
\begin{eqnarray}
Q_+(C)=  \sum_{C'} \pi_C(C') Q_+(C')
\label{eqqplus}
\end{eqnarray}
for all configurations $C$ different from the two ground states,
and the boundary conditions
\begin{eqnarray}
Q_+(C_+) && =1 \nonumber \\
Q_+(C_-) && =0
\label{bcqplus}
\end{eqnarray}

This suggests the following Ansatz for the antisymmetric $A(C)$ satisfying Eq. \ref{eqAe1zero} up to a normalization factor ${\cal N}$
\begin{eqnarray}
A^{ansatz}(C) ={\cal N}( 2 Q_+(C) -1 )= {\cal N} (1- 2 Q_-(C))
\label{acqplus}
\end{eqnarray}
using $Q_+(C)=Q_-(-C)$ one obtains $A(-C)=-A(C)$.

The only point where $Q_+(C)$ does not satisfy Eq. \ref{eqqplus}
are the two boundaries $C_+$ and $C_-$ where $Q_+$ is given by the b.c. 
corresponding to
\begin{eqnarray}
\frac{A^{ansatz}(C_+)}{{\cal N} }&& = 2 Q_+(C_+) -1 =1 \nonumber \\
\frac{A^{ansatz}(C_-)}{{\cal N} }&& = 2 Q_+(C_-) -1 =-1
\label{acqplusbc}
\end{eqnarray}

Let us now estimate $E_1$ for the Ansatz of Eq. \ref{acqplus} corresponding to
\begin{eqnarray}
\psi_1^{Ansatz}(C) = \psi_0(C) A^{ansatz}(C) ={\cal N}  \psi_0^{Ansatz}(C)( 2 Q_+(C) -1 )
\label{psi1qplus}
\end{eqnarray}
via
\begin{eqnarray}
E_1 && = \frac{< \psi_1^{Ansatz} \vert H_Q \vert \psi_1^{Ansatz} > }{< \psi_1^{Ansatz} \vert \psi_1^{Ansatz} >} \nonumber \\
&& = \frac{ \sum_C \psi_0^2(C) A^{ansatz}(C) \left[ W_{out}(C)  A^{ansatz}(C)
- \sum_{C'} W(C \to C')  A^{ansatz}(C')  \right] }
{\sum_C \psi_0^2(C) (A^{ansatz}(C))^2 }
\label{e1psi1qplus}
\end{eqnarray}
In the numerator, all configurations $C$ different from $C_+$ and $C_-$
give zero-contributions as a consequence of Eq. \ref{eqqplus}.
So the only contributions in the numerator come from $C=C_+$ and from $C=C_-$
where we may use the boundary conditions of Eq. \ref{acqplusbc}
to obtain
\begin{eqnarray}
E_1 
&& = \frac{  \psi_0^2(C_+)  \left[ W_{out}(C_+) 
- {\displaystyle \sum_{C'} } W(C_+ \to C') ( 2 Q_+(C') -1)  \right]
+  \psi_0^2(C_-)  \left[ W_{out}(C_-)  
- {\displaystyle \sum_{C'} } W(C_- \to C') (1- 2 Q_+(C')) \right] }
{\sum_C \psi_0^2(C) (2 Q_+(C) -1)^2 }
\nonumber \\
&&  = 2 \frac{  \psi_0^2(C_+) \left[ \sum_{C'} W(C_+ \to C') Q_-(C')  \right]
+  \psi_0^2(C_-) \left[  \sum_{C'} W(C_- \to C') Q_+(C') \right]  }
{\sum_C \psi_0^2(C) (2 Q_+(C) -1)^2 }
\label{e1psi1qplustwo}
\end{eqnarray}
The numerator involves the probability to reach first $C_-$ before returning to $C_+$
when one leaves $C_+$, and  the probability to reach first $C_-$ before returning to $C_+$
when one leaves $C_+$, which are the same by symmetry.

\subsection{ Application to the random ferromagnetic chain  near zero temperature }

We now focus on the random ferromagnetic chain of $N$ spins
of Eq \ref{Uspin1d}
with free boundary conditions for the two boundary spins $S_1$ and $S_N$.
Near zero temperature (Eq \ref{lowT}), we may neglect the configurations
 containing more than one domain-wall,
and work within the space of the following $(2N)$ configurations 
\begin{eqnarray}
\vert k>^{sym}_N && = \frac{1}{\sqrt 2} \left[ \vert S_1=..=S_k=-1; S_{k+1}=..=S_N=+1 > +\vert S_1=..=S_k=1; S_{k+1}=..=S_N=-1 >  \right]  \nonumber \\
\vert k>^{asym}_N && = \frac{1}{\sqrt 2} \left[ \vert S_1=..=S_k=-1; S_{k+1}=..=S_N=+1 > -\vert S_1=..=S_k=1; S_{k+1}=..=S_N=-1 >  \right]
\label{statesspinchain}
\end{eqnarray}
where $k=0,1,...,N-1$. In physical terms, $\vert 0>^{sym}$ and  $\vert 0 >^{asym}$ are the symmetric and antisymmetric
combination of the two ferromagnetic ground states where all spin have the same signs, whereas $\vert k sym>$ and  $\vert k asym>$ with $1 \leq k \leq N-1$ are the symmetric and antisymmetric
combination of the states where there exists a single domain-wall between the sites $(k,k+1)$.

We consider the quantum Hamiltonian 
\begin{eqnarray}
 {\cal H}_N 
&& =\sum_{2 \leq k \leq N-1 } 
G \left( J_{k-1,k} \sigma^z_{k-1} + J_{k,k+1} \sigma^z_{k+1} \right)
 \left[ e^{ - \beta \sigma^z_k \left(  J_{k-1,k} \sigma^z_{k-1}+ J_{k,k+1} \sigma^z_{k+1} \right) } - \sigma^x_k \right]
\nonumber \\
&& + G \left( J_{1,2}  \right)
\left[ e^{ - \beta \sigma^z_1  J_{1,2} \sigma^z_{2}} - \sigma^x_1 \right]
 + G \left( J_{N-1,N} \sigma^z_{k-1}  \right) 
\left[ e^{ - \beta \sigma^z_N  J_{N-1,N} \sigma^z_{N-1}} - \sigma^x_N \right]
\label{Hsimplechain}
\end{eqnarray}
where the amplitudes $G_k$ are given by a single even function $G(x)=G(-x)$ of the local field
(see Eq. \ref{Flocalfield}).

\subsubsection{ Two first eigenvectors within the single domain-wall approximation }

The ground state $\vert \psi_0>$ of zero energy is exactly known from Eq. \ref{psi0spins}
\begin{eqnarray}
\vert \psi_0 >_N =\frac{1}{Z_N} \sum_{S_1,...,S_N}   e^{ \frac{\beta}{2} \sum_{i=1}^{N-1} J_{i,i+1} S_i S_{i+1} } \vert S_1,...,S_N>
\label{psi0chainfull}
\end{eqnarray}
Within the reduced space of configurations containing no more than one domain-wall (Eq \ref{statesspinchain}),
the ground state reduces to
\begin{eqnarray}
\vert \psi_0 >_N \simeq \vert 0 >^{sym}_N + \sum_{k=1}^{N-1} e^{- \beta J_{k,k+1}} \vert k >^{sym}_N
\label{psi0chain}
\end{eqnarray}
near zero temperature

To respect the antisymmetry of Eq. \ref{psiunantisym},
 the first excited state will be a linear combination of the antisymmetric states of Eq. \ref{statesspinchain} 
\begin{eqnarray}
\vert \psi_1 >_N \simeq \vert 0 >^{asym}_N + \sum_{k=1}^{N-1} e^{- \beta J_{k,k+1}} A_N(k) \vert k >^{asym}_N
\label{psi1chain}
\end{eqnarray}
with some amplitudes $A_N(k)$ that we wish to determine.
The eigenvalue equation for this first excited state of the Hamiltonian of Eq. \ref{Hsimplechain} reads
\begin{eqnarray}
&& 0  =({\cal H}_N-E_1) \vert \psi_1 >_N  \label{eigenvaluepsi1}  \\
&& 
= \vert 0 >^{asym}_N \left[ -E_1 +  f^-_1 e^{ - \beta   J_{1,2}} \left(1-A_N(1) \right)+ f^-_N e^{ - \beta   J_{N-1,N}} \left(1+A_N(N-1) \right)  \right] \nonumber \\
&&
+ \vert 1 >^{asym}_N \left[ - E_1  e^{- \beta J_{1,2}} A_N(1)
+   f^-_1 \left(A_N(1)-1 \right)+ f^-_2 e^{ - \beta   J_{2,3}} \left(A_N(1)-A_N(2) \right)  \right]
 \nonumber \\
&& 
+ \sum_{k=2}^{N-2} \vert k >^{asym}_N \left[ - E_1  e^{- \beta J_{k,k+1}} A_N(k)
+  f^-_k e^{ - \beta   J_{k-1,k}} \left(A_N(k)-A_N(k-1) \right)+ f^-_{k+1} e^{ - \beta   J_{k+1,k+2}} \left(A_N(k)-A_N(k+1) \right)  \right]
 \nonumber \\
&& 
+  \vert N-1 >^{asym}_N \left[ - E_1  e^{- \beta J_{N-1,N}} A_N(N-1)
+  f^-_{N-1} e^{ - \beta   J_{N-2,N-1}} \left(A_N(N-1)-A_N(N-2) \right)+ f^-_N \left(A_N(N-1)+1 \right)  \right]
\nonumber
\end{eqnarray}
in terms of the numbers
\begin{eqnarray}
f^-_k \equiv G \left(J_{k-1,k} - J_{k,k+1} \right)
\label{gknumbers}
\end{eqnarray}

\subsubsection{ Ansatz with exit probabilities }

Instead of solving exactly the eigenvalue problem of an $N\times N$ matrix of Eq. \ref{eigenvaluepsi1},
we have proposed in section \ref{sec_ansatz} the following approximation :
in all coefficients involving $\vert k >^{asym}_N$ with $k=1,..,N-1$, we may neglect 
the term containing $E_1$ with respect to the others to obtain the $(N-1)$ equations for $k=1,..,N-1$
\begin{eqnarray}
f^-_k e^{ - \beta   J_{k-1,k}} \left(A_N^{ansatz}(k)-A_N^{ansatz}(k-1) \right)+ f^-_{k+1} e^{ - \beta   J_{k+1,k+2}} \left(A_N^{ansatz}(k)-A_N^{ansatz}(k+1) \right) =0
\label{recchain}
\end{eqnarray}
with the following boundary conditions
\begin{eqnarray}
A_N^{ansatz}(0)=1 \nonumber \\
A_N^{ansatz}(N)=-1
\label{bcchain}
\end{eqnarray}
The only remaining term in Eq \ref{eigenvaluepsi1} is then the first line involving $\vert 0 >^{asym}_N $
that determines the value of the energy $E_1$ as
\begin{eqnarray}
E_1^{ansatz}(N) = f^-_1 e^{ - \beta   J_{1,2}} \left(A_N^{ansatz}(0)-A_N^{ansatz}(1) \right)+ f^-_N e^{ - \beta   J_{N-1,N}} \left(A_N^{ansatz}(N-1)- A_N^{ansatz}(N)\right) 
\label{e1chain}
\end{eqnarray}
So we have replaced the eigenvalue problem of Eq. \ref{eigenvaluepsi1}
by a simpler homogeneous recurrence equation (Eq \ref{recchain})
with the boundary equations of Eq. \ref{bcchain}, that can be solved as follows.

\subsubsection{ Exact solution for exit probabilities in one dimension }

It is convenient to set as in Eq. \ref{acqplus}
\begin{eqnarray}
A^{ansatz}_N(k) = 2 Q_0(k) -1 =  1- 2 Q_N(k)
\label{acqpluschain}
\end{eqnarray}
where $Q_N(k)$ satisfies
\begin{eqnarray}
Q_N(k) = p_+(k) Q_N(k+1)+p_-(k) Q_N(k-1) 
\label{recchainq}
\end{eqnarray}
with the respective probabilities
\begin{eqnarray}
 p_+(k) && \equiv \frac{ f^-_{k+1} e^{ - \beta   J_{k+1,k+2}}}
{ f^-_{k+1} e^{ - \beta   J_{k+1,k+2}}+ f^-_k e^{ - \beta   J_{k-1,k}} }  \nonumber \\
 p_-(k) && \equiv \frac{ f^-_k e^{ - \beta   J_{k-1,k}}}
{ f^-_{k+1} e^{ - \beta   J_{k+1,k+2}}+ f^-_k e^{ - \beta   J_{k-1,k}} } =1-p_+(k)
\label{recchainqppm}
\end{eqnarray}
and the boundary conditions
\begin{eqnarray}
Q_N(0) && =0 \nonumber \\
Q_N(N) && =1
\label{qplusbc}
\end{eqnarray}
Then $Q_N(k)$ represents the probability to reach first the boundary $k=N$ rather than the boundary $k=0$
for a random walker starting at $k$ and moving with probabilities of Eq. \ref{recchainqppm}.
The well-known solution of this standard problem can be obtained by recurrence 
\cite{rec1d} using Kesten variables \cite{kestenv} and reads
\begin{eqnarray}
Q_N(k)  = \frac{R(1,k)}{R(1,N)}
\label{soluexitchain}
\end{eqnarray}
with 
\begin{eqnarray}
R(1,0) && =0 \nonumber \\
R(1,1) && =1 \nonumber \\
R(1,k \geq 2) && =1+\sum_{m=1}^{k-1} \prod_{n=1}^{m} \frac{ p_-(n)}{ p_+(n)}
\nonumber \\
R(1,N) && = 1+\sum_{m=1}^{N-1} \prod_{n=1}^{m} \frac{ p_-(n)}{ p_+(n)}
 = 1+ \frac{p_-(1)}{p_+(1)} + \frac{p_-(1)p_-(2)}{p_+(1)p_+(2)} + ... + \frac{p_-(1) p_-(2)...p_-(N-1)}{p_+(1) p_+(2)...p_+(N-1)}
\label{rsoluexitchain}
\end{eqnarray}

The corresponding estimate of the energy of Eq. \ref{e1chain} reads using Eq. \ref{acqpluschain}
\begin{eqnarray}
E_1^{ansatz}(N) && = f^-_1 e^{ - \beta   J_{1,2}} \left( (1-2 Q_N(0)) - (1-2 Q_N(1)) \right)
+ f^-_N e^{ - \beta   J_{N-1,N}} \left( 1-2 Q_{N}(N-1)  -(1-2 Q_N(N)) \right) 
\nonumber \\
&& = 2 f^-_1 e^{ - \beta   J_{1,2}} \left(  Q_N(1)-Q_N(0) \right)+ 2 f^-_N e^{ - \beta   J_{N-1,N}} \left(Q_N(N)-Q_N(N-1) \right)
\nonumber \\
&& =  2  \frac{ f^-_1 e^{ - \beta   J_{1,2}} R(1,1)  + f^-_N e^{ - \beta   J_{N-1,N}} \left[ R(1,N)-R(1,N-1)  \right] }{R(1,N)} 
\nonumber \\
&& =  2  \frac{ f^-_1 e^{ - \beta   J_{1,2}} + f^-_N e^{ - \beta   J_{N-1,N}} \left[ \frac{p_-(1) p_-(2)...p_-(N-1)}{p_+(1) p_+(2)...p_+(N-1)}  \right] }{  1+\sum_{m=1}^{N-1} \prod_{n=1}^{m} \frac{ p_-(n)}{ p_+(n)} } 
\label{e1chaininter}
\end{eqnarray}
in terms of the ratios (Eq \ref{recchainqppm})
\begin{eqnarray}
\frac{ p_-(k) }{ p_+(k)} && = \frac{ f^-_k e^{ - \beta   J_{k-1,k}}} { f^-_{k+1} e^{ - \beta   J_{k+1,k+2}}} 
 =\frac{f^-_k}{f^-_{k+1}} e^{  \beta ( J_{k+1,k+2} -  J_{k-1,k}) }
\label{ratiochainqppm}
\end{eqnarray}
Taking into account that absent links correspond to vanishing coupling $J_{0,1}=0=J_{N,N+1}$, 
one obtains
\begin{eqnarray}
\left[ \frac{p_-(1) p_-(2)...p_-(N-1)}{p_+(1) p_+(2)...p_+(N-1)}  \right] =\frac{f^-_1}{f^-_{N}}  e^{ - \beta  J_{1,2} + \beta  J_{N-1,N} }
\label{prod}
\end{eqnarray}
and
\begin{eqnarray}
R(1,N) && 
 = 1+ \frac{p_-(1)}{p_+(1)} + \frac{p_-(1)p_-(2)}{p_+(1)p_+(2)} + ... + \frac{p_-(1) p_-(2)...p_-(N-1)}{p_+(1) p_+(2)...p_+(N-1)}
\nonumber \\
&& = 1 + \frac{f^-_1}{f^-_{2}}  e^{  \beta  J_{2,3}  } 
+ \frac{f^-_1}{f^-_{3}} e^{ - \beta  J_{1,2} + \beta  J_{2,3}+ \beta  J_{3,4}  }
+ \frac{f^-_1}{f^-_{4}} e^{  - \beta  J_{1,2}+ \beta  J_{3,4} + \beta  J_{4,5} }
 + \frac{f^-_1}{f^-_{5}}  e^{ - \beta  J_{1,2} + \beta  J_{4,5}
+ \beta  J_{5,6} }  \nonumber \\
&&+ ... + \frac{f^-_1}{f^-_{N-1}} e^{ - \beta  J_{1,2} + \beta  J_{N-2,N-1} + \beta  J_{N-1,N} }
+ \frac{f^-_1}{f^-_{N}} e^{ - \beta  J_{1,2} + \beta  J_{N-1,N} }
\nonumber \\
&& = f^-_1 e^{ - \beta  J_{1,2}} \left[ \frac{e^{  \beta  J_{1,2}} }{f^-_1 } + \frac{e^{ \beta  J_{1,2}+ \beta  J_{2,3}  }}{f^-_{2}}+ \frac{ e^{  \beta  J_{2,3}+ \beta  J_{3,4}  } }{f^-_{3}} + \frac{e^{   \beta  J_{3,4} + \beta  J_{4,5} }}{f^-_{4}} 
+ ...  \frac{ e^{  \beta  J_{N-2,N-1} + \beta  J_{N-1,N} }}{f^-_{N-1}} 
+ \frac{e^{  \beta  J_{N-1,N} }}{f^-_{N}}  \right]
\nonumber \\
&& = f^-_1 e^{ - \beta  J_{1,2}} \sum_{k=1}^N \frac{ e^{  \beta  J_{k-1,k} + \beta  J_{k,k+1} }}{f^-_{k}} 
\label{denochain}
\end{eqnarray}
so that the energy of Eq. \ref{e1chaininter} reads
\begin{eqnarray}
&& E_1^{ansatz}(N) =
 2  \frac{ f^-_1 e^{ - \beta   J_{1,2}} + f^-_N e^{ - \beta   J_{N-1,N}} \left[ \frac{f^-_1}{f^-_{N}}  e^{ - \beta  J_{1,2} + \beta  J_{N-1,N} }\right]  }{  R(1,N) } 
 =    \frac{ 4 f^-_1 e^{ - \beta   J_{1,2}} }{ R(1,N) }  \nonumber \\
&& =    \frac{ 4  }{ \displaystyle \sum_{k=1}^N \frac{ e^{  \beta  J_{k-1,k} + \beta  J_{k,k+1} }}{f^-_{k}}  }
\label{e1chainfinal}
\end{eqnarray}
i.e. the equilibrium time reads using Eq \ref{gknumbers}
\begin{eqnarray}
t_{eq}(N) && = \frac{1}{ E_1^{ansatz}(N)}  =    \frac{1}{ 4  }
 \sum_{k=1}^N \frac{ e^{  \beta  J_{k-1,k} + \beta  J_{k,k+1} }}{f^-_{k}} 
= \frac{1}{ 4  }
  \sum_{k=1}^N \frac{ e^{  \beta  J_{k-1,k} + \beta  J_{k,k+1} }}{ G \left(J_{k-1,k} - J_{k,k+1} \right)} 
\label{teqchainfinal}
\end{eqnarray}
in agreement with the RG result of Eq. \ref{taueqgene} derived in the text.

\subsection{Exact renormalization rules in configuration space for escape probabilities $Q_{\pm}(C)$}

As a final remark, let us mention the link with previous works concerning renormalization rules {\it in configuration space}.
As explained in \cite{us_firstpassage}, backward master equations
satisfy exact renormalization rules in configuration space.
Upon the elimination of the configuration ${\cal C}_0$, 
 the surviving configurations 
${\cal C}$ satisfy the same equation as before (Eq \ref{eqqplus})
\begin{eqnarray}
W_{out}^{R}(C) Q_+(C)=  \sum_{C'} W^R(C \to C') Q_+(C')
\label{eqqplusR}
\end{eqnarray}
 where the renormalized transitions rates $W^R$ evolve with the RG equations
\begin{eqnarray}
W^{Rnew} \left({\cal C}\  \to  {\cal C}'  \right) && =  
W^R \left({\cal C}\  \to  {\cal C}'  \right)+ \frac{W^R \left({\cal C}\  \to  {\cal C}_0  \right) W^R \left({\cal C}_0\  \to  {\cal C}'  \right)}{W_{out}^R \left( {\cal C}_0 \right)} \nonumber \\
W_{out}^{Rnew} \left( {\cal C} \right) && =W_{out}^R \left( {\cal C} \right)
- \frac{W^R \left({\cal C}\  \to  {\cal C}_0  \right) W^R \left({\cal C}_0\  \to  {\cal C}  \right)}{W_{out}^R \left( {\cal C}_0 \right)}
 \label{rgrulesT}
\end{eqnarray}
These RG rules for backward master equations are exact and can be used \cite{us_firstpassage},
but only for small sizes as a consequence of the exponential growth of configurations.
The RG rules of Eq. \ref{rgrulesT} have been first derived via a Strong Disorder RG approach
\cite{us_rgconfig}.

Note that the RG rules of Eq. \ref{rgrulesT} can be rewritten directly for the renormalized
probabilities
\begin{eqnarray}
\pi^R_C(C') \equiv \frac{W^R(C \to C')}{W_{out}^{R}(C)}
\label{piR}
\end{eqnarray}
that evolve according to
\begin{eqnarray}
\pi^{Rnew}_C(C')   \equiv \frac{W^{Rnew}(C \to C')}{W_{out}^{Rnew}(C)}
= \frac{\pi^R_C(C') + \pi^R_C(C_0) \pi^R_{C_0}(C')}{1-\pi^R_C(C_0) \pi^R_{C_0}(C) }
\label{piRG}
\end{eqnarray}

\section{ Dependence on the choice of the dynamics   }

\label{sec_diff}

In this Appendix, we describe how the equilibrium time depends on the choice of the dynamics satisfying detailed balance

\subsection{ Case of the Glauber dynamics }

For the Glauber dynamics, one expects that the dynamical barrier coincides with the
maximal energy cost on the optimal path between the two ground states.
For instance for the one-dimensional random ferromagnetic chain,
the result of Eq. \ref{taueqglauber} satisfies
 \begin{eqnarray}
\frac{1}{\beta} \ln \left[  t^{Glauber}_{eq}(N) \right]
 &&  = {\max \limits_{0 \leq k \leq N-1} } \ \left( 2 J_{k,k+1} \right) 
 = {\max \limits_{0 \leq k \leq N} } \ \left(U_{N}^{(k,N-k)} -  U_{N}^{GS} \right) 
\label{linkbarrierworst}
\end{eqnarray}
where $U_{N}^{(k,N-k)} $ represents the energy of the configuration where the 
 first $k$ spins are $(-1)$, whereas all others spins are $(+1)$.

To better understand the differences with the simple dynamics described below,
it is useful to write the result of Eq. \ref{taueqglauber} for the two smallest sizes,
with $N=2$ and $N=3$ spins
\begin{eqnarray}
t_{eq}^{glauber}(N=2) &&  =  \frac{1}{ 2 }\left[ 1 +   e^{2 \beta J_{1,2}  }  \right]
\nonumber \\
 t_{eq}^{glauber}(N=3) &&  =  \frac{1}{ 2 }\left[ 1 +   e^{2 \beta J_{1,2}} +e^{2 \beta J_{2,3}}   \right]
\label{taueqglaubersmall}
\end{eqnarray}

\subsection{ Case of the simple dynamics  }

For the 'simple' dynamics, the correspondence of Eq. \ref{linkbarrierworst}
between the dynamical barrier and the maximal energy cost of a single domain wall 
does not hold, as can be seen already for 
the one-dimensional case with $N=2$ and $N=3$ spins 
since Eq \ref{taueqsimple} reads 
\begin{eqnarray}
t_{eq}^{simple}(N=2) && = \frac{1}{ 2 }    e^{ \beta J_{1,2} } 
\nonumber \\
t_{eq}^{simple}(N=3) && = \frac{1}{ 4 }  \left[ 
 e^{ \beta J_{1,2} }+ e^{ \beta (J_{1,2}+J_{2,3}) }+  e^{ \beta  J_{2,3} } \right]
\label{taueqsimplesmall}
\end{eqnarray}
For $N=2$, the difference by a factor of $2$ between the dynamical barriers
can be understood from the differences between the transitions rates
for the simple dynamics 
(Eq \ref{Wsimple})
 \begin{eqnarray}
W^{simple}(++ \to +-) && =W^{simple}(++ \to -+) = e^{- \beta J_{1,2}} 
\nonumber \\
W^{simple}( +- \to ++ ) && =W^{simple}(+- \to --) = e^{+ \beta J_{1,2}} 
\label{rates2spinssimple}
\end{eqnarray}
and for the Glauber dynamics (eq \ref{glauber})
 \begin{eqnarray}
W^{Glauber}(++ \to +-) && =W^{Glauber}(++ \to -+) = \frac{e^{- \beta J_{1,2}} }{e^{+ \beta J_{1,2}}+e^{- \beta J_{1,2}} }
 = \frac{e^{- 2 \beta J_{1,2}} }{1+e^{- 2 \beta J_{1,2}} }
\nonumber \\
W^{Glauber}( +- \to ++ ) && =W^{Glauber}(+- \to --) =  \frac{e^{+ \beta J_{1,2}} }{e^{+ \beta J_{1,2}}+e^{- \beta J_{1,2}} }
 = \frac{1 }{1+e^{- 2 \beta J_{1,2}} }
\label{rates2spinsglauber}
\end{eqnarray}
For $N=2$ spins, the equilibrium time is determined by the rate $W(++ \to +-)$
to create a domain-wall when starting from one ground state
(the time to eliminate the domain-wall is then negligible), and these two rates are 
respectively of order $ e^{- \beta J_{1,2}}  $ for the simple dynamics and of order $e^{- 2 \beta J_{1,2}}$ for the Glauber dynamics.
One could argue that the Glauber dynamics is more 'physical', in the sense that all transitions rates remain bounded
near zero-temperature, whereas in the 'simple' dynamics transition rates corresponding to a decrease of the energy
diverge near zero temperature.
Nevertheless, one expects on physical grounds that
 the difference between the dynamical barriers of the two dynamics
remains of order $O(1)$, as found in this article for the one-dimensional case and for the tree case,
and as found in the companion paper \cite{us_dyndyson} for the Dyson hierarchical Ising model.

\end{document}